\documentstyle[12pt]{article}

\newcommand{\newsection}[1]{\section{#1}\setcounter{equation}{0}}
\def\fcskp{\baselineskip 12pt}              
\newcommand{\fc}{\footnotesize \fcskp \advance\itemsep by -6pt}
\def\DREZ{\hbox{${\rm DR\overline{EZ}}$}}
\def\DREZpr{\hbox{${\rm DR\overline{EZ}}'$}}
\def\DRED{{\rm DRED}}
\def\NDR{{\rm NDR}}
\def\NDRpr{\hbox{${\rm NDR}'$}}
\def\HV{{\rm HV}}
\def\PV{{\rm PV}}
\def\CONT{{\em cont}}
\def\LATT{{\em lat}}

\def\TRUE{{\em true}}
\def\PEN{{\em pen}}
\def\SUB{{\em sub}}

\def\tC{{\widetilde C}}
\def\tQ{{\widetilde Q}}
\def\half{{\textstyle{1\over2}}}
\def\sixth{{\textstyle{1\over6}}}
\def\frac#1#2{{\textstyle{#1\over#2}}}

\def\Tr{\mathop{\rm Tr}\nolimits}
\def\Dsl{\,\raise.15ex\hbox{/}\mkern-13.5mu D}
\def\bar#1{\overline{#1}}
\def\chibar{\bar\chi}
\def\qbar{{\bar{q}}}
\def\ubar{{\bar{u}}}

\def\sbar{{\bar{s}}}
\def\vev#1{\langle #1 \rangle}

\def\cb{{\cal B}}

\def\cg{{\cal G}}

\def\cm{{\cal M}}
\def\co{{\cal O}}
\def\cp{{\cal P}}

\def\vdag{{\vphantom{\dag}}}
\def\g#1{\gamma_{#1}}

\def\pl{(1\!+\!\g5)}
\def\pr{(1\!-\!\g5)}
\def\gam#1{\overline{(\gamma_{#1}\otimes I)} }
\def\ggam#1{\overline{\overline{(\gamma_{#1}\otimes I)} } }
\def\ixi#1{\overline{(I\otimes\xi_{#1})} }
\def\iden{\overline{(I\otimes I) } }

\def\sfno#1#2{\overline{(\gamma_{#1}\otimes\xi_{#2})}}

\def\sf#1#2#3#4{\overline{(\gamma_{#1}\otimes\xi_{#2})}_{#3#4}}
\def\sfdag#1#2#3#4{%
\overline{(\gamma^{\dag}_{#1}\otimes\xi^{\dag}_{#2})}_{#3#4}}
\def\ssf#1#2#3#4{\overline{\overline{(\gamma_{#1}\otimes\xi_{#2})}}_{#3#4}}
\def\opold#1#2{\,[#1\times#2]}
\def\opoldc#1#2#3{\,[#1\times#2]_{#3}}
\def\opnew#1#2#3{\,[#1_{#3}\times#2_{#3}]}

\def\opergi#1{(\g{#1}\otimes I)}
\def\opergx#1#2{(\g{#1}\otimes\xi_{#2})}
\def\bfco{{\stackrel{\longrightarrow}{\co}}}
\def\bfcp{{\stackrel{\longrightarrow}{\cp}}}

\begin{document}
\include{psfig}
\begin{titlepage}
 \null
 \begin{center}
 \makebox[\textwidth][r]{UW/PT-93-1}
 \par\vspace{0.25in} 
  {\Large
	PERTURBATIVE CORRECTIONS FOR STAGGERED \\ [0.5em]
        FOUR-FERMION OPERATORS}
  \par
 \vskip 2.0em 
 {\large 
  \begin{tabular}[t]{c}
	Stephen R. Sharpe \footnotemark\\[1.em]
	\em Physics Department, FM-15, University of Washington \\
	\em Seattle, WA 98195 \\[1.5em]
	Apoorva Patel \footnotemark\\[1.em]
	\em CTS and SERC, Indian Institute of Science\\
	\em Bangalore, 560012, India\\
  \end{tabular}}
 \par \vskip 3.0em
 {\large\bf Abstract}
\end{center}
\quotation
We present results for one-loop matching coefficients between
continuum four-fermion operators, defined in the Naive Dimensional
Regularization scheme, and staggered fermion operators of various types.
We calculate diagrams involving gluon exchange between quark lines,
and ``penguin'' diagrams containing quark loops.
For the former we use Landau gauge operators, 
with and without $O(a)$ improvement,
and including the tadpole improvement suggested by Lepage and Mackenzie.
For the latter we use gauge-invariant operators.
Combined with existing results for two-loop
anomalous dimension matrices and one-loop matching coefficients,
our results allow a lattice calculation of the
amplitudes for $K\bar K$ mixing and $K\to\pi\pi$ decays
with all corrections of $O(g^2)$ included.
We also discuss the mixing of $\Delta S=1$ operators with lower
dimension operators, and show that, with staggered fermions,
only a single lower dimension operator
need be removed by non-perturbative subtraction.
\endquotation

\footnotetext[1]{Email: sharpe@galileo.phys.washington.edu}
\footnotetext{Email: adpatel@cts.iisc.ernet.in}
\vfill
\mbox{October 1993}
\end{titlepage}

\newsection{INTRODUCTION}
\label{sintro}

Many of the low-energy effects of the electroweak interactions can
be expressed as matrix elements of four-fermion operators
between hadronic states. For example, CP-violation in $K-\bar{K}$ mixing
is proportional to such a matrix element (parameterized by $B_K$).
Knowledge of these matrix elements is required to extract the parameters
of the electroweak theory.
Since the matrix elements involve non-perturbative physics,
lattice QCD is well suited for their calculation.\footnote{%
For a recent review using lattice matrix elements to constrain 
standard model parameters see Ref. \cite{lusignoli}.}\
A necessary ingredient, however, is the relationship 
between continuum and lattice four-fermion operators.
This ``matching'' can be done using perturbation theory,
since the operators differ only for momenta close to the lattice cut-off.

The general form of the relationship between lattice and continuum 
four-fermion operators is
\begin{equation}
\label{contlatt}
\co^{\CONT}_i = \co^{\LATT}_i  + {g^2\over 16\pi^2} 
\sum_j \left(\gamma_{ij}^{(0)} \ln(\pi/\mu a) + c_{ij}\right)  \co^{\LATT}_j
+ O(g^4) + O(a)\ ,
\end{equation}
where $\mu$ is the renormalization scale used to define the 
continuum operators, and $a$ is the lattice spacing.
In other words, to calculate the matrix elements of a continuum operator 
$\co^\CONT_i$, we must use the linear combination of lattice operators 
given by the right-hand-side of this equation.
$\gamma_{ij}^{(0)}$ is the one-loop anomalous dimension matrix
(using the conventions of Ref. \cite{buras}), and $c_{ij}$ is the
finite part of the one-loop matching coefficient.
The finite part depends on the continuum regularization scheme,
on the type of lattice fermions used, and on the precise definition of
the lattice operators.

In this paper we calculate $c_{ij}$ for two classes of
lattice four-fermion operators composed of staggered fermions.
In general, such operators involve quark and antiquark fields at
different positions, and must be made gauge-invariant.
Our first calculation concerns operators which are defined by fixing
to Landau gauge and dropping all gauge link factors. 
We compute the one-loop matching coefficients for such ``Landau-gauge'' 
operators coming from diagrams in which a gluon is exchanged 
between quark lines (see Fig. \ref{4fermionfig}), 
which we refer to as X diagrams.
These are the only contributions for operators transforming as
{\bf 27}'s under flavor SU(3), 
e.g. the $\Delta S=2$ operator leading to $K-\bar{K}$ mixing,
 and the $\Delta I=3/2$ part of the $K\to\pi\pi$ amplitude.
We consider both the simplest Landau-gauge operators
and ``smeared'' operators having no $O(a)$ corrections at tree-level. 
These operators have been used extensively in numerical
simulations \cite{bkprl,sharpelat91,fukugitalat91,ishiprl93},
and the results have small enough statistical errors that it is 
important to include the perturbative corrections.

Our second calculation is of the contribution of ``penguin'' diagrams
to the matching coefficients (see Fig. \ref{penguinfig}). 
At one-loop, the contributions of X diagrams and penguin diagrams add,
$c_{ij} = c_{ij}^X + c_{ij}^\PEN$, and can be treated separately.
Penguin diagrams are present for
operators transforming as octets under flavor SU(3).
Such operators contribute to the $\Delta I=1/2$ part of the $K\to\pi\pi$
amplitude, and in particular to CP-violation in these decays.
Numerical calculations of their matrix elements are possible \cite{epsilonpus},
although less advanced than those for ${\bf 27}$'s.
A technical problem with the calculations is that 
there is mixing not only with four-fermion operators but also 
with bilinears of lower dimension.
These must be subtracted non-perturbatively, e.g.
using the method of Ref. \cite{cps}.
If one uses operators made gauge-invariant by the 
inclusion of gauge links (``gauge-invariant'' operators),
then the mixing is constrained by gauge invariance.
In contrast, Landau-gauge four-fermion operators can mix with 
gauge non-invariant bilinears, and require many more subtractions.
For this reason, we present results 
for penguin diagrams only for gauge-invariant operators.

In the continuum, the mixing of octet operators with bilinears of 
lower dimension is restricted by chiral symmetry and CP. 
It turns out that only a single such bilinear can appear, 
with just the overall factor undetermined.
This result is crucial for the subtraction method of Ref. \cite{cps} to work.
Complications arise with lattice operators, however,
because of the fermion doubling problem. 
The issue is well understood for Wilson fermions, 
where the breaking of chiral symmetry allows mixing with
a number of lower dimension bilinears \cite{nonpertref,cpskpipi,sigmawilson},
necessitating the use of more elaborate subtraction schemes.
With staggered fermions, on the other hand,
mixing is restricted by the axial $U(1)$ symmetry.
While a large number of lower dimension bilinears can appear, 
only one of these contributes to physical matrix elements.
This allows one to use the subtraction method of Ref. \cite{cps}.
The fact that only one operator need be subtracted 
has been either implicitly or explicitly assumed in previous
discussions \cite{toolkit,wius,book},
but is demonstrated here.

The mixing coefficients $c_{ij}$ depend on the choice of
scheme used to define the continuum operators.
In physical matrix elements this dependence cancels with that
of the two-loop anomalous dimension matrix.
Thus for our results to be useful they must be given for a
continuum scheme in which the two-loop anomalous dimension has been
calculated. Results for both {\bf 27} and {\bf 8} operators are
available in both Naive Dimensional Regularization (\NDR)
and 't Hooft-Veltman (\HV---Ref. \cite{thooft}) schemes. 
We choose to give results for the NDR scheme.

This work extends previous studies using staggered fermions.
Gauge-invariant four-fermion operators were discussed 
by Daniel and Sheard \cite{daniel}, and the complete results for some
of these operators were given by Sheard \cite{sheard}.
The calculation for Landau-gauge four-fermion operators turns out to be 
simpler than that for gauge-invariant operators,
requiring only the matching coefficients for bilinears.
These have been given in our previous work \cite{patelsharpe},
henceforth referred to as PS.
Ishizuka and Shizawa have independently calculated the 
corrections from X diagrams for unsmeared Landau-gauge four-fermion operators 
\cite{ishiprl93,tsukubapert},
and our results are in complete agreement with theirs.
This is a non-trivial check since the calculations are done using somewhat
different methods.
They have also calculated the corrections from X diagrams
for gauge-invariant four-fermion operators.
No previous calculations have studied penguin diagrams.
The complete one-loop results for gauge-invariant SU(3) octets
can thus be obtained by combining $c_{ij}^X$ from Ref. \cite{tsukubapert}
with those for $c_{ij}^\PEN$ given here.

A single staggered fermion corresponds to $N_f=4$ Dirac fermions
in the continuum limit.
We follow Ref. \cite{toolkit} and introduce a staggered
fermion ``species'' for each continuum flavor, i.e. $\chi_u$, $\chi_d$,
$\chi_s$ etc. 
Thus the continuum theory to which we match 
has $N_f$ degenerate copies of each quark in QCD.
It is straightforward to relate the matrix elements in this continuum 
theory to those in continuum QCD~\cite{toolkit,wius,book}. 

The outline of this paper is as follows. In sect. \ref{snotation}
we explain our notation for four-fermion operators, 
and discuss their symmetries.
Section \ref{sxdiagrams} gives the details of the computations of
the X diagrams for Landau-gauge operators,
and sect. \ref{sxresults} the corresponding results for operators
of phenomenological interest.
In sect. \ref{spenguin} we present the calculation of the
penguin diagrams, and give results for the matching of the
continuum operators which contribute to kaon decays.
We close in sect. \ref{smixing} with a discussion of the mixing
with lower dimension operators.
Two appendices collect results on Fierz transformations and the
matching coefficients for bilinears.

Preliminary accounts of this work have been given
in Refs. \cite{book} and \cite{ssringberg}.
Some of the numerical results in these two works are wrong,
and are corrected here.

\newsection{NOTATION AND DEFINITIONS}
\label{snotation}

In this section we set up the notation required to specify the 
four-fermion operators, and discuss their symmetries. While this section is
self-contained, it draws heavily on the presentation given in PS,
which should be consulted for further details.

\subsection{Bilinears}

We use a gamma matrix basis for both spin and flavor matrices.
To enumerate this basis we use ``hypercube vectors'',
four-vectors whose components are 0 or 1,
which are combined using modulo-2 operations.
A general gamma matrix is labeled by a hypercube vector
\begin{equation}
  \g{S} = \g{1}^{S_1} \g{2}^{S_2} \g{3}^{S_3} \g{4}^{S_4} \ .
\end{equation}
We use Euclidean space gamma matrices, which are hermitian, and
satisfy $\{\g\mu,\g\nu\}=2\delta_{\mu\nu}$.
An alternative basis is built out of the complex conjugate matrices 
$\xi_\mu=\g\mu^*$,
\begin{equation}
  \xi_F = \xi_1^{F_1} \xi_2^{F_2} \xi_3^{F_3} \xi_4^{F_4} \ ,
\end{equation}
where $F$ is another hypercube vector.
To keep the notation clear, we always use $\xi$ matrices for flavor,
and $\gamma$ matrices for spin.
Useful shorthand notations are
\begin{equation}
S=\widehat\mu \Rightarrow \g{S}=\g\mu \ ,\quad
S=(1,1,1,1)  \Rightarrow \g{S}=\g5 \ ,\quad
S=(0,0,0,0)  \Rightarrow \g{S}=I \ ,
\end{equation}
\begin{equation}
  \g{S}  \g{S'}   = \g{SS'}\ , \quad
  \xi_{S}\xi_{S'} = \xi_{SS'} \ .
\end{equation}

We label quark fields in a continuum theory with four degenerate
fermions using upper case letters, e.g. $Q_{\alpha,a}$. 
Here $\alpha$ is the spinor index, and $a$ the flavor index,
both running from 1 to 4.
For the moment we suppress color indices.
A general bilinear is specified by a spin and a flavor matrix
\begin{equation}
  \co_{SF}^\CONT =
  \bar Q_{\alpha,a} \g{S}^{\alpha\beta} \xi_F^{ab} Q_{\beta,b} 
  \ = \     \bar Q_{\alpha,a} \opergx{S}{F}^{\alpha a,\beta b} Q_{\beta,b} 
  \ \equiv\ \bar Q \opergx{S}{F} Q \ .
\end{equation}
In the second step we combine spin and flavor matrices into a single
$16\times16$ matrix $\opergx SF$.

We use a variety of lattice transcriptions of these continuum operators.
All make use of a set of matrices which are related to $\opergx SF$
by a unitary transformation \cite{daniel,kieu} :
\begin{eqnarray}
  \sf S F A B &\equiv&
  \frac14 \Tr[ \g{A}^{\dag} \g{S}^{\vdag} \g{B}^{\vdag} \g{F}^{\dag} ] \\
  \label{hypercuberep}
  &=&\sum_{\alpha,a,\beta,b}      (\frac12\xi_{A})^{\alpha a}
  \opergx{S}{F}^{\alpha a,\beta b} (\frac12\g{B})^{\beta b} \ .
\end{eqnarray}
In the new basis we have traded the indices $\{ \alpha,a \}$ 
for a hypercube vector $A$.
Perturbative calculations in momentum space require another unitarily
equivalent set of matrices 
\begin{equation}
  \label{polerep}
  \ssf S F A B \equiv
  \sum_{CD} \frac14 (-)^{A.C}\ \sf S F C D \ \frac14 (-)^{D.B} \ .
\end{equation}
All three sets of matrices satisfy the multiplication rule exemplified by
\begin{equation}
\label{multiplymatrices}
  \sum_B \sf{S}{F}AB\ \sf{S'}{F'}BC = \sf{SS'}{FF'}AC \ .
\end{equation}

To define lattice bilinears we divide the lattice into 
$2^4$ hypercubes in one of the 16 possible ways.
Points on the original lattice are then specified by a vector $y$
labeling the hypercubes (with all components even),
and a hypercube vector $C$ determining the position within the hypercube.
The 16 components of the staggered fermion field $\chi$ for a given $y$ 
are collected in to a single hypercube field
\begin{equation}
  \chi(y)_C = \frac14 \chi(y+C) \ .
\end{equation}
In the continuum limit this becomes equal to $Q$,
when expressed in the appropriate basis \cite{kluberg}
\begin{equation}
  \chi(y)_C \longrightarrow Q(y)_C = (\half\xi_C)^{\beta b} Q_{\beta,b} \ .
\end{equation}
Thus the lattice operator
\begin{equation}
\label{klubergeq}
  \co_{SF}(y) = \sum_{C,D}\ \chibar(y)_C\ \sf SFCD \ \chi(y)_D 
  \equiv \chibar \sfno SF \chi \ .
\end{equation}
has the same flavor, spin and normalization
as $\co_{SF}^\CONT=\bar Q \opergx SF Q$ in the continuum limit.

An important property of a bilinear is its ``distance'',
\begin{equation}
\label{deltaeqn}
\Delta=\sum_{\mu=1}^{4} |S_\mu - F_\mu|^2 \ .
\end{equation}
For the lattice bilinear of Eq. \ref{klubergeq} this is the number of
links between the quark and antiquark field.
{}From this we construct the ``distance-parity'' of the bilinear,
$(-1)^\Delta$, which is $+1$ if the distance is even, $-1$ if it is odd.
This parity is useful because of the $U(1)_A$ symmetry, valid when $ma=0$,
\begin{equation}
\chi \to \exp\left[ i \alpha \sfno55 \right] \chi \ ,\quad
\chibar \to  \chibar \exp\left[ i \alpha \sfno55 \right] \ ,
\end{equation}
where $\alpha$ is the rotation angle.
Under this transformation, even distance bilinears rotate
\begin{equation}
\label{evenrotation}
(-)^\Delta= 1\ :\quad
\chibar \sfno{S}{F} \chi \to \cos\alpha\ \chibar \sfno{S}{F} \chi
+ i \sin\alpha\ \chibar \sfno{S5}{F5} \chi \ ,
\end{equation}
while odd distance bilinears are unchanged.
It follows that even and odd distance bilinears cannot mix, and that the
renormalization of even distance bilinears related by multiplication
by $\sfno55$ is identical \cite{smitvink}.
It has been noticed in Ref. \cite{daniel} and in PS that, at one-loop,
the latter result holds also for odd distance bilinears, 
although this does not follow from $U(1)_A$ symmetry.
We now understand this result. As long as we consider diagrams in which
the $\chibar$ and $\chi$ fields are not contracted, 
then we can do separate $U(1)_A$ rotations on the two fields.
In particular, if the rotation angles $\alpha$ for $\chi$ and $\chibar$
are opposite, then odd distance bilinears rotate as in
Eq. \ref{evenrotation}, while even distance bilinears are unchanged.
It follows that odd distance bilinears related by multiplication by
$\sfno55$ will renormalize identically.

As mentioned in the introduction,
we consider both gauge-invariant and Landau-gauge operators.
In the former, we insert between $\chibar$ and $\chi$ a matrix
given by the average of the products of gauge links 
along the shortest paths connecting the quark and antiquark.
In the latter, we fix to Landau gauge,
and then use Eq. \ref{klubergeq} without any gauge links.
Landau-gauge operators have the advantages of being simple
to implement numerically,
and of having Feynman rules with no coupling to gluons.
Their disadvantage is that, nonperturbatively, there may be Gribov copies,
in which case the definition of the operators is ambiguous.
In perturbation theory there is, however, no ambiguity, and we do not
consider this problem further here.

As shown in PS, both gauge-invariant and Landau-gauge bilinears
differ from the corresponding continuum bilinears at $O(a)$.
For Landau-gauge bilinears, however, 
it is straightforward to ``improve''
the operators so that they have only $O(a^2)$ corrections 
at tree level \cite{book}.
One simply replaces the quark field in Eq. \ref{klubergeq} by
\begin{equation}
  \chi(y)_A \to \frac14 \sum_\mu \chi(y+ 2\hat\mu[1-2A_\mu] )_A =
           \frac1{16} \sum_\mu \chi(y+A+ 2\hat\mu[1-2A_\mu] ) \ ,
\end{equation}
and performs a similar replacement for the antiquark field.
The resulting ``smeared'' operators are spread over a $4^4$ hypercube.
Since the fields are shifted by an even number of lattice spacings,
smearing does not affect the spin-flavor content
or the distance-parity of the operator.

We also consider a further improvement of the various operators. 
Lepage and Mackenzie have shown that a large part of the 
difference between lattice and continuum operators is due to
fluctuations of the gauge links, and in particular due to tadpole diagrams 
\cite{lepagemackenzie}.
These corrections can be partially summed using
a rescaled gauge link $U'_\mu= U_\mu/u_0$ in both the action and the
operators.
As shown in PS, the prescription for staggered fermion bilinears
is to multiply them by $u_0^{1-n_U}$ in order to match more closely 
with the continuum operators.
Here $n_U$ is the number of gauge links in the operator
($n_U=0$ for Landau-gauge operators).
We call operators including this factor ``tadpole-improved''.
There are various choices for $u_0$;
for reasons given in PS we choose $u_0=\langle\frac13 \Tr(U)_L\rangle$,
the normalized trace of the link in Landau gauge.

\subsection{Four-fermion operators}

The general four-fermion operator is specified by two spins and two flavors
\begin{equation}
\label{fourfermieq}
\chibar \sfno{S'}{F'} \chi \ \chibar\sfno{S}{F}\chi
\equiv \sfno {S'}{F'} \ \sfno {S}{F} \ .
\end{equation}
The second form is a useful abbreviation for the operator which we only
use when there is no danger of confusion with the product of two single-barred
matrices such as that in Eq. \ref{multiplymatrices}.
The two bilinears in Eq. \ref{fourfermieq} can be of any type 
(gauge-invariant, smeared Landau-gauge, etc.),
although we only use operators in which both are of the same type.
For a given type of bilinear, there are $256^2$ four-fermion operators,
so that the matching matrices are huge: $65536\times65536$!
Fortunately we need only calculate small blocks of this matrix.
In particular, as explained below, matrix elements of phenomenological
interest can be obtained using operators for which $S=S'$ and $F=F'$, 
which we refer to as ``diagonal'' operators,
and the associated operators $\sfno S F\ \sfno{S5}{F5}$.

For certain calculations we can restrict our attention further,
and consider only the 35 linear combinations of the diagonal operators which
transform as the identity representation of the Euclidean lattice rotation
group.\footnote{%
Among the 65536 operators of Eq. \ref{fourfermieq}
there are many other operators transforming as the identity representation,
but they are off-diagonal, i.e. $S'\ne S$ and/or $F'\ne F$
 \cite{sheard}.}
Such ``diagonal scalar'' operators suffice,
for example, in a calculation of $B_K$ \cite{bkprl,book}.
There are few enough such operators that
we can display some explicit results, 
rather than simply giving numerical answers.

To enumerate the 35 diagonal scalar operators we use a compact notation.
Since the spin and flavor of both bilinears are the same, 
we need only specify the spin and flavor of one bilinear.
The first 25 operators are those in which the spin and flavor
independently form scalars. For the spin we use the basis
\[
S = I\cdot I\ ,\qquad 
V = \sum_\mu \gamma_\mu \cdot \gamma_\mu\ ,\quad
T = \sum_{\mu<\nu} \gamma_\mu\gamma_\nu \cdot \gamma_\nu\gamma_\mu\ ,
\]
\[
A = \sum_\mu \gamma_\mu\gamma_5 \cdot \gamma_5\gamma_\mu\ ,\quad
P = \g5 \cdot \g5\ ,
\]
where the dot separates the matrices in the two bilinears.\footnote{%
We are using $S$ to denote both a hypercube vector representing all 
possible spins, and a scalar operator.
Where confusion might arise, we denote the scalar operator by $I$.}\
Note that the spin matrix in the second bilinear is the hermitian
conjugate of that in the first. 
We use exactly the same basis for flavor except that $\gamma\to\xi$.
The first 25 scalar operators are obtained by combining 
the five possible spins with the five flavors.
Our notation is exemplified by
\begin{eqnarray}
\opold AP &=& \sum_\mu \sfno{\mu5}{5}\ \sfno{5\mu} 5\ , \\
\opold VT &=& 
\sum_\mu \sum_{\nu<\rho} \sfno{\mu}{\nu\rho}\ \sfno{\mu}{\rho\nu} \ .
\end{eqnarray}
In the continuum limit, these operators become Lorentz scalars.

In the remaining 10 operators the spin and flavor are not independent.
We denote these operators as
\begin{equation}
\opnew VV\mu = \sum_\mu \sfno \mu \mu\ \sfno \mu \mu \ ,
\end{equation}
with $\opnew VA\mu$, $\opnew AV\mu$ and $\opnew AA\mu$ defined similarly;
\begin{equation}
\opnew VT\mu  = \sum_{\scriptstyle \mu,\nu \atop \scriptstyle \mu\ne\nu}
\sfno \mu{\mu\nu}\  \sfno \mu{\nu\mu} -
\sfno \mu{\mu\nu5}\ \sfno \mu{5\nu\mu}  \ ,
\end{equation}
and the analogous $\opnew T V \mu$;
\begin{equation}
\opnew AT\mu  = \sum_{\scriptstyle \mu,\nu \atop \scriptstyle \mu\ne\nu}
\sfno {\mu5}{\mu\nu5}\  \sfno {5\mu}{5\nu\mu} -
\sfno {\mu5}{\mu\nu}\ \sfno {5\mu}{\nu\mu}  \ ,
\end{equation}
and the analogous $\opnew T A \mu$;
and finally
\begin{equation}
\opnew T T -  = \sum_{\mu<\nu}
\sfno {\mu\nu}  {\mu\nu}\  \sfno {\nu\mu}  {\nu\mu}  -
\sfno {\mu\nu}  {\mu\nu5}\ \sfno {\nu\mu}  {5\nu\mu} 
\end{equation}
with $\opnew T T +$ defined similarly.\footnote{%
The definitions of $\opnew TT\pm$ 
differ from those used in Ref.~\cite{book} by a factor of two.}\
Linear combinations of these operators are written 
\begin{equation}
\opnew{(V+A)} T \mu = \opnew V T \mu + \opnew A T \mu \ .
\end{equation}
These 10 operators are lattice scalars, but do not become
Lorentz scalars in the continuum limit.

Each four-fermion operator comes in two ``color-types''.
The color indices can  be contracted either between bilinears 
or within each bilinear
\begin{equation}
\co_I  = \sfno{S}{F}_{ab}\sfno{S}{F}_{ba}\ ;\ 
\co_{II} = \sfno{S}{F}_{aa}\sfno{S}{F}_{bb}\ .
\end{equation}
This notation indicates the number of color loops when 
each of the bilinears is
contracted with an external color singlet meson:
operators of type (I) give rise to a single color trace, 
while those of type (II) produce two traces.

Lastly, we consider the implications of the $U(1)_A$ symmetry for the
mixing of four-fermion operators.
For diagrams in which the fields in the operator are
not contracted with each other (i.e. X diagrams), we can do
independent $U(1)_A$ rotations on each of the four fields.
Several results follow from this.
\begin{itemize}
\item
The distance-parity of each bilinear cannot be changed by mixing.
Diagonal operators can be classified as even or odd distance 
because both bilinears have the same distance-parity.
Thus it follows that even and odd distance diagonal 
four-fermion operators cannot mix.
\item
The mixing coefficients for the operators $\sfno SF\ \sfno{S5}{F5}$ 
can be obtained from those for $\sfno SF\ \sfno SF$ by applying
an appropriate axial rotation to the second bilinear.
This is equivalent to multiplying the second bilinear by $\sfno55$
from either the right or left.
\item
The corrections are also unchanged if we multiply both bilinears by $\sfno55$
from, say, the left. We call this operation $\cp$.
This invariance is derived by applying an 
appropriate axial rotation to both bilinears.
\end{itemize}
These results are valid to all orders in perturbation theory,
up to corrections of $O(ma)$. We have checked them explicitly in
our one-loop calculations.
They are not valid for penguin diagrams, 
as we discuss in more detail in sects. \ref{spenguin} and \ref{smixing}.

\newsection{CALCULATING X DIAGRAMS}
\label{sxdiagrams}

\subsection{Overview of method}

In this section we describe the calculation of the 
contribution to one-loop matching coefficients from X diagrams.
We do this only for Landau-gauge operators.
As mentioned in the introduction, we first match operators in QCD
onto those in a continuum theory with $N_f$ degenerate versions of each quark.
We consider the following operators in the latter theory
\begin{equation}
\label{27conteqn}
\co_i^\CONT = 
\overline{S} \opergx{S}{F} D \ \overline{S} \opergx{S}{F} D \ .
\end{equation}
This choice of flavors puts the operator into a {\bf 27} of flavor SU(3),
and ensures that only X diagrams contribute to the matching calculation.
This operator will match onto lattice operators having the same flavor
\begin{equation}
\label{27eqn}
\co_j =
\chibar_s \sfno {S'}{F'} \chi_d \ \chibar_s \sfno {S''}{F''}\chi_d \ .
\end{equation}
Throughout this section, and the following, we assume that such a choice 
of flavors has been made, but do not show the flavor indices explicitly.

We consider only diagonal operators in Eq. \ref{27conteqn},
because the continuum operators that we are interested in are diagonal in spin,
and because actual numerical calculations take matrix elements between states
having the same flavor $F$.
There is no technical difficulty in extending the calculations
to off-diagonal operators.
The diagonal continuum operators do, however, match onto lattice operators
which are both diagonal and off-diagonal, as indicated in Eq. \ref{27eqn}.
In particular, at one-loop, operators appear having $F'\ne F''$,
although $S'=S''$. The matrix elements of such operators vanish, however,
between lattice states of the same flavor, up to corrections 
suppressed either by powers of $a$, or by additional factors of $g^2$.
Thus we need only consider matching with diagonal lattice operators.

We are also interested in the matching matrices for odd parity operators
of the form
\begin{equation}
\overline{S} \opergx{S}{F} D \ \overline{S} \opergx{S5}{F5} D \ .
\end{equation}
As explained in the previous section, however,
we can obtain the matching coefficients for these with no extra work.

The relation between continuum and lattice operators,
Eq. \ref{contlatt}, is derived by comparing the matrix elements 
of lattice and continuum operators between external states consisting
of two quarks and two antiquarks.
The coefficients $\gamma_{ij}^{(0)}$ and $c_{ij}$ do not depend 
on the external momenta or quark masses
except through terms suppressed by powers of the lattice spacing.
Thus we set the external momenta and the quark masses to zero. 
Infrared divergences are regulated by adding a gluon mass,
which can be set to zero in the final result.
Ultraviolet divergences are removed in the continuum 
using a variety of renormalization schemes.

\begin{figure}
\centerline{\psfig{file=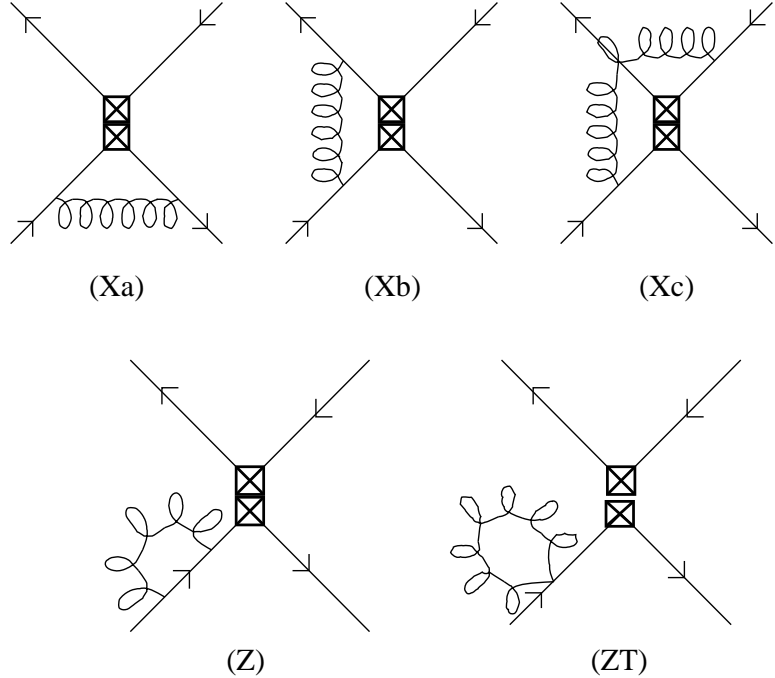,height=3.5truein}}
\caption[4fermionfig]{\fc
Diagrams contributing to renormalization of four-fermion operators
for Landau-gauge operators.
The two boxes represent the two bilinears which make up the operators.
Only half of the X diagrams are shown;
in the others the gluon exchanges are between the remaining two
fermion lines.
Similarly there are additional Z and ZT diagrams.
}
\label{4fermionfig}
\end{figure}

The diagrams which contribute at one-loop are
those of Fig. \ref{4fermionfig}, 
with the gluon propagator being in Landau gauge.
Following Refs. \cite{daniel} and PS, we call the first three 
Xa, Xb and Xc diagrams,
and the remaining two Z and ZT diagrams.
For tadpole-improved operators the contribution of ZT diagram 
cancels with the $O(g^2)$ term in the expansion of $u_0^{-1}$,
so one can exclude these diagrams from the start.
This is an example of how the tadpole-improved operators are
similar to continuum operators, for the latter are corrected by
all except the ZT diagrams.

We first consider the one-loop lattice matrix element.
We take the operator to be of color-type (II), 
and explain at the end how the results are
modified for color-type (I) operators.

The Xa diagrams combine with the self-energy corrections Z and ZT
just as in the calculation for bilinears. 
The corrections can thus be obtained from the results for
Landau-gauge bilinears. One must simply add the results for
each of the two bilinears contained in the four-fermion operator.
We discuss this in more detail in Appendix \ref{appcorr}.

It turns out that one can also obtain the results
for the Xb and Xc diagrams using only those for bilinears.
This is done in a number of steps.
First, we note that to bring the Xc diagrams into canonical form
one of the quark propagators has to be conjugated,
since the gluon connects two quarks (or antiquarks).
The fermion-gluon vertex remains unaffected,
but conjugation reverses the momentum flow along the fermion
propagator and thus changes its sign when $m=0$.
Z and ZT diagrams, which are unaffected by conjugation,
can now be added to the Xb diagrams, and subtracted from the
Xc diagrams, such that each calculation is like that for a bilinear.
Schematically
\begin{equation}
Xb+Xc = Xb - Xc^{\dag} = (Xb + Z + ZT) - (Xc + Z + ZT)^{\dag} \ .
\end{equation}
Note that these manipulations are possible because the color factors 
are the same for the Xb and Xc diagrams.

For the Xb diagrams, the next step is to do a simultaneous
Fierz transformation on the spin and flavor indices.
This is necessary because the gluon connects a quark and antiquark
which are part of different bilinears in the original form of the operator.
The Fierz transformations for spin (or flavor) alone are standard.
We use the notation
\begin{equation}
\label{fierzeqn}
\Gamma^{\vdag}_{\alpha\beta} \Gamma^{\dag}_{\gamma\delta} =
\sum_{\Gamma'} f(\Gamma,\Gamma') \Gamma'^{\vdag}_{\alpha\delta} 
\Gamma'^{\dag}_{\gamma\beta} \ ,
\end{equation}
where $\Gamma$ is a general Dirac matrix.
Combining two such transformations, we find
\begin{equation}
\sf {S} {F} A B \sfdag {S} {F} C D 
= \sum_{S',F'} f(S,S') f(F,F')
\sf {S'}{F'} A D  \sfdag {S'}{F'} B C \ .
\end{equation}
For the diagonal scalars, which Fierz transform into one another,
we give the Fierz table in Appendix \ref{appfierz}.
In general we use a computer program to calculate the Fierz table.
Having Fierz transformed, we can use the results for the bilinear
corrections, and then Fierz transform back to the original basis.

For the Xc diagrams, an additional step is required. 
First, one of the bilinears is conjugated using
\begin{equation}
\sf {S} {F} A B = \sfdag {S} {F} B A \ .
\end{equation}
One then does a Fierz transformation, calculates the correction,
Fierz transforms back, and finally conjugates again.
All these steps are straightforward, but, given the number of operators
involved, they rapidly lead to an explosion of terms. In practice we 
have written a computer program to do the various steps,
and have checked it by working out some examples by hand.

The same method works for the four-fermion operators of color type (I).
The only difference is that the Z and ZT diagrams combine with the 
Xb diagrams to form a bilinear correction, while one must add and subtract
Z and ZT diagrams to the Xa and Xc diagrams respectively.

For continuum operators the steps in the calculation are the same,
except for one subtlety which we return to shortly.
The end result is that lattice and continuum corrections combine 
just as in the calculation for bilinears. 
Thus what we require from our previous work is the 
bilinear matching coefficients defined by
\begin{equation}
  \label{bilinres}
  \cb^{\CONT}_i = \cb^{\LATT}_i \left( 1 + C_F {g^2\over 16\pi^2} 
  [ 2(1-\sigma_S) \ln(\pi/\mu a) + c_i] \right) + \hbox{\rm off-diagonal}
  \ .
\end{equation}
Here $\cb$ denotes a bilinear, $C_F=4/3$ is the color factor,
and $\sigma_S$, which depends only on the spin of the bilinear,
is $(4,1,0,1,4)$ for spin tensors $(S,V,T,A,P)$.
The coefficients $c_i$ depend upon the spin and flavor matrices
in the bilinear, upon its type (e.g. smeared {\em vs.} unsmeared),
and upon the continuum renormalization scheme.
Numerical values are given in PS.

We do not need to keep the off-diagonal components of
the one-loop mixing matrix in Eq. \ref{bilinres}. 
These cause mixing between operators having the same spin
but different flavor.
For example, the continuum vector current $\bar Q \opergx\mu\nu Q$ 
($\mu\ne\nu$) matches onto a linear combination of
the lattice operators $\chibar \sfno\mu\nu \chi$ 
and $\chibar \sfno\mu\mu \chi$ (no sum on $\mu$).
Such matching coefficients imply that continuum four-fermion 
operators constructed of two bilinears with the same flavor and spin 
will match in part with lattice operators in which the spins of
the two bilinears are the same but the flavors differ.
As discussed above, the contribution of such operators can be 
dropped in the matrix elements between states having the same flavor.
We note, however, that in a two-loop perturbative calculation one must
retain such off-diagonal mixing at intermediate stages.

We now return to the subtlety concerning continuum regularization schemes.
Fierz transformation and charge conjugation apply only if the gamma matrices
are in 4-dimensions, as in Dimensional Reduction (\DRED) and Pauli-Villars
(\PV) schemes. For schemes involving n-dimensional gamma matrices, 
such as \NDR, the method described above will not work. 
In such schemes, one has to enlarge the basis of operators to include
``evanescent'' operators which vanish when restricted to 4-dimensions.
There is considerable ambiguity in how one defines evanescent operators,
and different choices lead to different results for the finite parts
of diagrams, and thus for the coefficients $c_i$.
The method described above amounts to a particular, and unconventional,
choice of evanescent operators. We denote the result with this choice
as being in the \NDRpr\ scheme. To relate the results to the conventional
choice of evanescent operators of Refs. \cite{buras,newburas,martinelli},
we must make a finite correction.
We explain this further in subsect. \ref{sintondr}.

In the ``correction'' step of the above calculations
we need to convert the corrections for bilinears 
into those for four-fermion operators, keeping only the diagonal part.
For a general diagonal four-fermion operator of the form of
Eq. \ref{fourfermieq}, we simply sum the corrections 
to each of the constituent bilinears. 
In practice we implement this numerically.
For the 35 diagonal scalars, which are linear combinations of operators
receiving different corrections,
the results are worked out in Appendix \ref{appcorr}.

\subsection{Color factors}
\label{scolorfactors}

Each gluon exchange gives rise to a color factor
\begin{equation}
\label{colorfierz}
\sum_{\alpha=1}^{8} T^\alpha_{ab} T^\alpha_{cd} = 
\frac12 \delta_{ad}\delta_{cb} - \frac16 \delta_{ab}\delta_{cd} \ .
\end{equation}
In general this mixes operators of the two color types.
Since the color mixing does not depend on the spin or flavor of the operator,
it can be factored out.
To display the result we use a vector notation
\begin{equation}
\bfco_i =
\left( \begin{array}{c} \co_{i(I)}\\ \co_{i(II)} \end{array} \right)
\ ,
\end{equation}
where $i$ only labels the spin and flavor of the operator.
The matching equation can then be written
\begin{equation}
\label{cmdefeq}
\bfco^\CONT_i = \bfco^\LATT_i
+ {g^2\over 16\pi^2} \sum_j \left( 
  {\cm}^a_{ij} \stackrel{\longleftrightarrow}{C_a} 
+ {\cm}^b_{ij} \stackrel{\longleftrightarrow}{C_b} 
+ {\cm}^c_{ij} \stackrel{\longleftrightarrow}{C_c}  \right) 
\bfco^\LATT_j \ ,
\end{equation}
where $\cm^{a,b,c}_{ij}$ are the contributions from the Xa,b,c diagrams,
respectively, together with associated self-energy diagrams,
with color factors (and $g^2/16\pi^2$) removed.
$C_{a,b,c}$ are the corresponding color matrices
\begin{equation}
 \stackrel{\longleftrightarrow}{C_a} 
	  = \frac16 \left( \begin{array}{rr} 
                   -1 & 3 \\ 0 & 8 \end{array} 
            \right)
\ ,\quad
 \stackrel{\longleftrightarrow}{C_b} 
	  = \frac16 \left( \begin{array}{rr} 
                   8 & 0 \\ 3 & -1 \end{array} 
            \right)
\ ,\quad
 \stackrel{\longleftrightarrow}{C_c}
          = \frac16 \left( \begin{array}{rr} 
                   -1 & 3 \\ 3 & -1 \end{array} 
            \right)
\ .
\end{equation}

\subsection{An example: $\opold VS$}
\label{ssexample}
To explain the method in more detail we 
calculate the lattice operator which
matches onto the continuum operator $\opold VS$. 
This is part of the matching calculation needed to extract $B_K$.
For simplicity of presentation,
we adopt the following notation for rows of the matching matrices
\begin{equation}
\label{corrnotation}
\cm^a(\opold VS) \equiv  
\left(\sum_j \cm^a_{ij} \co_j \right)_{i=\opold VS} \ .
\end{equation}

\subsubsection{Xa diagrams}
For the Xa diagrams, we can take the result directly from appendix
\ref{appcorr}
\begin{equation}
\nonumber
\cm^a(\opold VS) = 2 c_{VS} \opold VS \ .
\end{equation}

\subsubsection{Xb diagrams}
We must perform the operations Fierz-correct-Fierz.
Using the appendices we find
\begin{eqnarray*}
\lefteqn{16 \cm^b(\opold VS) 
= (c_{SS}-c_{SP}+3c_{VV2}+c_{VV0}-3c_{VA2}-c_{VA4})
(\opold SV + \opold PA)} \\
&& \mbox{} + (-2c_{SV}+2c_{SA}-2c_{VS}+2c_{VP}) 
(\opold SV - \opold PA) \\
&& \mbox{} + (c_{SS}-c_{SP}-3c_{VV2}-c_{VV0}+3c_{VA2}+c_{VA4})
(\opold SA + \opold PV) \\
&& \mbox{} + (2c_{SV}-2c_{SA}-2c_{VS}+2c_{VP}) 
(\opold SA - \opold PV) \\
&& \mbox{} + (c_{SS}+6c_{ST}+c_{SP}+3c_{VV2}+c_{VV0}+3c_{VA2}+c_{VA4})
(\opold VS + \opold AP) \\
&& \mbox{} + (4c_{SV}+4c_{SA}+ c_{VS}+3c_{VT1}+3c_{VT3}+ c_{VP})
(\opold VS - \opold AP) \\
&& \mbox{} + (c_{SS}-2c_{ST}+c_{SP})
\opold {(V+A)}T \\
&& \mbox{} + (c_{VS}- c_{VT1}- c_{VT3} + c_{VP}) 
\opold {(V-A)}T \\
&& \mbox{} + (c_{SS}+6c_{ST}+c_{SP}-3c_{VV2}-c_{VV0}-3c_{VA2}-c_{VA4})
(\opold VP + \opold AS) \\
&& \mbox{} + (-4c_{SV}-4c_{SA}+ c_{VS}+3c_{VT1}+3c_{VT3}+ c_{VP})
(\opold VP - \opold AS) \\
&& \mbox{} + (c_{SS}-c_{SP})
\opold T{(V+A)} \\
&& \mbox{} + (-2c_{SV}+2c_{SA})
\opold T{(V-A)} \\
&& \mbox{} + (c_{VV0}-c_{VV2}+c_{VA4}-c_{VA2})
\opnew {(V+A)} T \mu  \\
&& \mbox{} + (c_{VV0}-c_{VV2}-c_{VA4}+c_{VA2})
\opnew T {(V+A)} \mu \\
&& \mbox{} + (2c_{VT3}-2c_{VT1})
\opnew T {(V-A)} \mu \\
&& \mbox{} + \hbox{\rm off-diagonal}
\end{eqnarray*}
We have simplified the answer by setting $\mu=\pi/a$, so that the
anomalous dimension factor is absent. It is simple to reinstate this
factor, however, as it depends only on the spin of the bilinear 
being corrected (Eq. \ref{bilinres}),
and thus can be determined by replacing $c_{SF}$ with
$2(1-\sigma_S)\ln(\pi/\mu a)$ in the above equation.
For the present example, this gives
\begin{equation}
\cm^b(\opold VS; {\rm anom.\ dim.}) 
=  -6 \ln(\pi/\mu a)(\opold VS + \opold AS) \ .
\end{equation}
Note that no even-distance operators appear in $\cm_b$, 
as required by the axial symmetries.

\subsubsection{Xc diagrams}

The first step is to conjugate, which for $\opold VS$ has no effect.
Thus the ensuing Fierz-correct-Fierz steps give the same result
as for the Xb diagrams. Finally, we must conjugate again, and multiply
the total result by $-1$. This gives
\begin{eqnarray*}
\lefteqn{- 16 \cm^c(\opold VS) 
= (c_{SS}-c_{SP}+3c_{VV2}+c_{VV0}-3c_{VA2}-c_{VA4})
(\opold SV - \opold PA) } \\
&& \mbox{} + (-2c_{SV}+2c_{SA}-2c_{VS}+2c_{VP}) 
(\opold SV + \opold PA) \\
&& \mbox{} - (c_{SS}-c_{SP}-3c_{VV2}-c_{VV0}+3c_{VA2}+c_{VA4})
(\opold SA - \opold PV) \\
&& \mbox{} - (2c_{SV}-2c_{SA}-2c_{VS}+2c_{VP}) 
(\opold SA + \opold PV) \\
&& \mbox{} + (c_{SS}+6c_{ST}+c_{SP}+3c_{VV2}+c_{VV0}+3c_{VA2}+c_{VA4})
(\opold VS - \opold AP) \\
&& \mbox{} + (4c_{SV}+4c_{SA}+ c_{VS}+3c_{VT1}+3c_{VT3}+ c_{VP})
(\opold VS + \opold AP) \\
&& \mbox{} - (c_{SS}-2c_{ST}+c_{SP})
\opold {(V-A)}T \\
&& \mbox{} - (c_{VS}- c_{VT1}- c_{VT3} + c_{VP}) 
\opold {(V+A)}T \\
&& \mbox{} + (c_{SS}+6c_{ST}+c_{SP}-3c_{VV2}-c_{VV0}-3c_{VA2}-c_{VA4})
(\opold VP - \opold AS) \\
&& \mbox{} + (-4c_{SV}-4c_{SA}+ c_{VS}+3c_{VT1}+3c_{VT3}+ c_{VP})
(\opold VP + \opold AS) \\
&& \mbox{} - (c_{SS}-c_{SP})
\opold T{(V-A)} \\
&& \mbox{} - (-2c_{SV}+2c_{SA})
\opold T{(V+A)} \\
&& \mbox{} - (c_{VV0}-c_{VV2}+c_{VA4}-c_{VA2})
\opnew {(V-A)} T \mu  \\
&& \mbox{} - (c_{VV0}-c_{VV2}-c_{VA4}+c_{VA2})
\opnew T {(V-A)} \mu \\
&& \mbox{} - (2c_{VT3}-2c_{VT1})
\opnew T {(V+A)} \mu \\
&& \mbox{} + \hbox{\rm off-diagonal}
\end{eqnarray*}
The anomalous dimension term can be reconstructed from this result
in the same way as for $\cm_b$
\begin{equation}
\cm^c(\opold VS; {\rm anom.\ dim.}) 
=  6 \ln(\pi/\mu a)(\opold VS - \opold AS) \ .
\end{equation}
Once again, only odd-parity operators appear.

\newsection{RESULTS FOR X DIAGRAMS}
\label{sxresults}

\subsection{General features}

Using the method explained in the previous section it is straightforward
to obtain the $256\times 256$ block of the matching matrices $\cm^{a,b,c}$ 
connecting diagonal four-fermion operators to each other.
Both the Fierz transformations and the (diagonal part of) 
the bilinear corrections can be represented as $256\times256$ matrices.
To obtain numerical results, we have written a computer
program to multiply these matrices appropriately and put in the color factors.
We have done this both for the full matrix and for the $35\times35$
block connecting diagonal scalar operators to each other.
We have checked these programs by comparing them to each other,
and to various analytical results such
as those of the previous section.

We do not need the full matching matrix if we take matrix elements 
between states of definite external flavor. 
For example, in the mixing of $\opold VS$, we are only interested 
in the matching coefficients for the operators with flavor $S$.
Operators with other flavors will contribute only at $O(g^2 a)$
or $O(g^4)$.
Because odd-distance operators mix only with each other, 
it suffices to consider the mixing of $\opold VS$ with $\opold AS$.
Thus we want only a $2\times2$ block of the matrices $\cm^{a,b,c}$.
Including color types, we then have a $4\times4$ matching matrix.
We give results only for such sub-matrices.

We quote results for two types of operator: unsmeared and smeared,
both tadpole improved.
The results for operators without tadpole improvement are the same 
except that diagonal matching matrix elements are shifted:
$c_{ii} \to c_{ii}-24.4661$. This simple change follows from the fact
that tadpoles appear only on external lines.

To present numerical results we must choose a continuum regularization 
scheme. We want to use \NDR, but
as explained above, the method of the previous section leads to
results in a slightly different scheme, \NDRpr.
Since we have to add a finite correction to convert from \NDRpr\ to \NDR,
we might as well use an intermediate scheme for which the methods
of the previous section apply, 
i.e. a scheme involving 4-dimensional gamma matrices. 
We choose Dimensional Reduction \cite{siegel} 
using the ``easy subtraction'' scheme of Ref. \cite{uclapert} (\DREZpr).
The prime indicates that, for four-fermion operators, our method differs
slightly from that of Ref. \cite{uclapert}.
Another virtue of using \DREZpr\ is that the numerical results for bilinears
are given in PS using this scheme. This allows the diligent 
reader to check the results given here more easily.

It is straightforward to convert from \DREZpr\ to other schemes.
Changing schemes introduces a spin dependent, but flavor
independent shift in the bilinear matching coefficients
\begin{equation}
  \label{mixingcs}
  c_i({\rm scheme\ 2}) = c_i({\rm scheme\ 1}) 
	+ t_S({\rm scheme\ 2}) - t_S({\rm scheme\ 1}) \ .
\end{equation}
For the schemes we consider $t_I=t_P$ and $t_V=t_A$.
The values of $t_S$, for \DREZpr, 
dimensional reduction using the subtraction scheme of Ref. \cite{altarelli}
(\DRED),
naive dimensional regularization with an anticommuting $\gamma_5$ (\NDRpr),
and Pauli-Villars using the subtraction scheme of Ref. \cite{daniel}
(\PV) \footnote{%
For the \PV\ scheme we use $M_g= \mu \sqrt{e}$ for the gluon regulator mass.},
are
\begin{equation}
\label{tsequation}
t_S = 
\left\{
\begin{array}{rrrrll} 
(& 0.5,&  0,&0.5)&\qquad\qquad
				&(\DREZpr) \\ 
(& 0.5,&0.5,&0.5)&	&(\DRED) \\ %
(&-0.5,&  0,&1.5)&	&(\NDRpr)  \\ 
(& 0,  &  0,&  0)&	&(\PV)   \\ %
  \end{array} \right.
\end{equation}
for spin tensors $(I,V,T)$, respectively.
These results are taken from PS, except that we have added the result for
the \PV\ scheme.
The shift in $c_i$ introduces a change in $\cm^{a,b,c}$ which depends
on the spins of the four-fermion operators but not on their flavor.
Including color factors one finds the change in the matching matrices to be
\begin{equation}
\label{schemeshifting}
  c_{ij}({\rm scheme\ 2}) = c_{ij}({\rm scheme\ 1}) 
  + \sum_{S=I,V,T} d^S_{ij} 
   \left[t_S({\rm scheme\ 2})- t_S({\rm scheme\ 1})\right] \ ,
\end{equation}
where the three matrices $d^{I,V,T}$ are independent of flavor.
We quote results for these below, in addition to those for $c_{ij}$.

A change in renormalization scale $\mu$ is a special case of changing
schemes. Using Eq. \ref{bilinres} it is simple to show that
\begin{equation}
\label{anomdimdefeqn}
\gamma_{ij}^{(0)} = -6 d^I_{ij} + 2 d^T_{ij} \ .
\end{equation}

We now present results for various operators that are used in
present calculations. Following PS, to give some idea of the size
of the corrections we set $\mu=\pi/a$ and use a boosted coupling $g^2=1.8$.
With these choices, the anomalous dimension contribution to matching vanishes,
and the matching matrix is to be multiplied by $g^2/16\pi^2\approx 1/90$.

\subsection{$\opold VS$ and $\opold AS$}
\label{VSASsubsect}

This is the part of the matching matrix needed for a calculation of $B_K$
using external pseudo-Goldstone pions.
We are interested in the submatrix of $c_{ij}$ with indices
\begin{equation}
\label{VSindices}
i,j = \left( \opoldc VSI, \opoldc VS{II}, \opoldc ASI, \opoldc AS{II}\right)
\ . 
\end{equation}
Using symmetry operation $\cp$ (defined in the previous section),
this matrix is the same as that with indices
\begin{equation}
\label{APindices}
i,j = -\left( \opoldc API, \opoldc AP{II}, \opoldc VPI, \opoldc VP{II}\right)
 \ .
\end{equation}
The minus sign arises because, under $\cp$,
$V\leftrightarrow -A$.
(The other transformations are $S\leftrightarrow P$ and $T\leftrightarrow T$.)
The sign is, however, irrelevant here, 
since it is an overall sign for the operators and does not affect $c_{ij}$.

The matrices needed for changing schemes are
\begin{equation}
d^I= \frac16 
\left(\begin{array}{rrrr} 
 9 & -3 &  7 &  3 \\
 0 &  0 &  6 & -2 \\
 7 &  3 &  9 & -3 \\
 6 & -2 &  0 &  0 
\end{array}\right) 
\ , \qquad
d^V= \frac16 
\left(\begin{array}{rrrr} 
 7 &  3 & -7 & -3 \\
 0 & 16 & -6 &  2 \\
-7 & -3 &  7 &  3 \\
-6 &  2 &  0 & 16 
\end{array}\right) \ , \qquad
d^T=0 \ ,
\end{equation}
so that $ \gamma^{(0)} = -6 d^I$.
For the finite part of the coefficients, the results
for the (tadpole-improved) Landau-gauge operators are\footnote{%
The accuracy of these and following results is
$\approx \pm 0.0001$.}
\begin{eqnarray}
\label{VSASres}
c(\hbox{\rm unsmeared}) &=&
\left(\begin{array}{rrrr} 
 2.0450 & -0.6817 & 6.8467 &  2.9343 \\
 0      &  0      & 5.8686 & -1.9562 \\
 6.8467 &  2.9343 & 2.5726 & -2.2643 \\
 5.8686 & -1.9562 & 0      & -4.2203 
\end{array}\right) 
\ , \\
\label{VSASsmres}
c(\hbox{\rm smeared}) &=&
\left(\begin{array}{rrrr} 
 10.0757 & -4.6508 & 12.5283 &  5.3693 \\
  0      & -3.8768 & 10.7385 & -3.5795 \\
 12.5283 &\ 5.3693 & 10.2457 & -5.1609 \\
 10.7385 & -3.5795 &  0      & -5.2371 
\end{array}\right) \ .
\end{eqnarray}
These corrections are quite small---multiplying by $1/90$ one sees that
they are 10\% or less. Without tadpole improvement the corrections
would be somewhat larger.

Ishizuka and Shizawa have reported results for a variety of
unsmeared operators, both Landau-gauge and gauge-invariant \cite{tsukubapert}.
They quote results
for a rectangular block of the mixing matrix for Landau-gauge operators:
$c_{ik}$, with $i$ given by Eq. \ref{VSindices},
and $k$ running over all operators.
This contains the square matrix Eq. \ref{VSASres}.
Using the conversion formula
\begin{equation}
- c_{ij}^\LATT(\hbox{Ref. \cite{tsukubapert}})
= c_{ij} + (\ln\pi + \frac7{12}) \gamma^{(0)}_{ij} \ ,
\end{equation}
and accounting for our opposite definition of $A$ and $T$,
we find that the results agree within the quoted errors.

\subsection{$\opold SS$, $\opold TS$ and $\opold PS$}

For calculations of $\epsilon'$, matrix elements of the operators
$\opold SS$, $\opold PS$, $\opold PP$ and $\opold SP$ are required.
The matching coefficients for these operators also involve $\opold TS$
and $\opold TP$.
Thus we present results for the submatrices having indices
\begin{equation}
\label{SSindices}
i,j = \left(  \opoldc SSI, \opoldc SS{II}, \opoldc TSI, \opoldc TS{II},
	      \opoldc PSI, \opoldc PS{II} \right) \ ,
\end{equation}
which are equal to the submatrices having indices
\begin{equation}
\label{PPindices}
i,j = \left(  \opoldc PPI, \opoldc PP{II}, \opoldc TPI, \opoldc TP{II},
	      \opoldc SPI, \opoldc SP{II} \right) \ .
\end{equation}
For the anomalous dimension parts we find
\begin{eqnarray}
d^I &=& \frac1{24}
\left(\begin{array}{rrrrrr} 
 1 & 21 &  7 &  3 &  9 & -3 \\
 0 & 64 &  6 & -2 &  0 &  0 \\
42 & 18 & 54 &-18 & 42 & 18 \\
36 &-12 &  0 &  0 & 36 &-12 \\
 9 & -3 &  7 &  3 &  1 & 21 \\
 0 &  0 &  6 & -2 &  0 & 64
\end{array}\right) 
\ , \\
d^V &=& \frac12
\left(\begin{array}{rrrrrr} 
 3 &-1 & 0 & 0 &-3 & 1 \\
 0 & 0 & 0 & 0 & 0 & 0 \\
 0 & 0 & 0 & 0 & 0 & 0 \\
 0 & 0 & 0 & 0 & 0 & 0 \\
-3 & 1 & 0 & 0 & 3 &-1 \\
 0 & 0 & 0 & 0 & 0 & 0 
\end{array}\right) \ , \\
d^T &=& \frac1{24}
\left(\begin{array}{rrrrrr} 
 27 & -9 & -7 & -3 & 27 & -9 \\
  0 &  0 & -6 &  2 &  0 &  0 \\
-42 &-18 & 10 & 18 &-42 &-18 \\
-36 & 12 &  0 & 64 &-36 & 12 \\
 27 & -9 & -7 & -3 & 27 & -9 \\
  0 &  0 & -6 &  2 &  0 &  0
\end{array}\right) 
\ ,
\end{eqnarray}
while the finite parts are
\begin{equation}
\label{SSres}
\left(\begin{array}{rrrrrr} 
  1.3864 & -20.1799 & 2.0878 &   0.8948 &   0.75   &  -0.25 \\
  0      & -59.1533 & 1.7895 &  -0.5965 &   0      &   0    \\
 12.5267 &   5.3686 & 6.7342 &  -5.5910 &  12.5267 &   5.3686 \\
 10.7371 &  -3.5791 & 0      & -10.7371 & -10.7371 &  -3.5791 \\
  0.75   &  -0.25   & 2.0878 &   0.8948 &  -9.3253 &  11.9549 \\
  0      &   0      & 1.7895 &  -0.5965 &   0      &  26.5395
\end{array}\right) \ , 
\end{equation}
\begin{equation}
\label{SSsmres}
\left(\begin{array}{rrrrrr}
 -5.9359 &   2.4296 &  3.9817 &   1.7064 &   0.75   &  -0.25 \\
  0      &   1.3531 &  3.4128 &  -1.1376 &   0      &   0    \\
 23.8899 &  10.2385 & 17.1791 & -10.3096 &  23.8899 &  10.2385 \\
 20.4771 &  -6.8257 &  0      & -13.7499 &  20.4771 &  -6.8257 \\
  0.75   &  -0.25   &  3.9817 &   1.7064 &  -9.2877 &  12.4850 \\
  0      &   0      &  3.4128 &  -1.1376 &   0      &  28.1675 \\
\end{array}\right) 
\end{equation}
respectively for unsmeared and smeared operators.

These corrections are larger than those for the vector and axial operators.
For the unsmeared operators, the range is roughly $-0.65$ to $0.3$,
which is sufficiently large that it may not be possible to do reliable
computations.
Calculations with smeared operators should be more reliable,
since the corrections range only from $-0.15$ to $0.3$.

\subsection{Flavor $\xi_{35}$ and $\xi_{345}$ operators}

Numerical calculations have also been performed with 
lattice pions having flavors other than $\xi_5$. For example,
Ref. \cite{hmks} uses flavors $\xi_{35}$, $\xi_{45}$ and $\xi_{345}$.
The continuum operators that are required, e.g.
\begin{equation}
\label{xi35opeq}
\sum_\mu \sfno{\mu5}{35} \ \sfno{5\mu}{53} \ ,
\end{equation}
are diagonal but not scalar. Thus we must use the
full basis of the 256 diagonal operators.
We present results only for flavors $\xi_{35}$ and $\xi_{345}$,
as one can extract the results for all other flavors
using the $\cp$ operation and the lattice rotation symmetry,

In this subsection we give results only for the $c_{ij}$.
The $d^S$ matrices, which are independent of flavor, 
have already been given above. The only subtle point is that one must
reexpress them in the different bases that we use here.

We begin with the results for the vector and axial operators needed for $B_K$.
For external flavor $\xi_{35}$ states, the continuum operators we need are
\begin{equation}
i = [\sum_\mu\sfno{\mu5}{35} \ \sfno{5\mu}{53}_{(I,II)},
     \sum_\mu\sfno{\mu}{35} \ \sfno{\mu}{53}_{(I,II)}] \ .
\end{equation}
to match onto which we need the lattice operators
\begin{eqnarray*}
j&=&[\sum_{\mu\ne3}\sfno{\mu5}{35} \ \sfno{5\mu}{53}_{(I,II)},
     \sum_{\mu\ne3}\sfno{\mu}{35} \ \sfno{\mu}{53}_{(I,II)},\\
&&\quad
     \sfno{35}{35} \ \sfno{53}{53}_{(I,II)},
     \sfno{3}{35} \ \sfno{3}{53}_{(I,II)}] \ .
\end{eqnarray*}
The matching matrix for the unsmeared operators is
\begin{equation}
\left(\begin{array}{rrrrrrrr} 
  2.6784& -2.5817&  6.8467&  2.9343&  3.9134& -6.2867&  6.8467&  2.9343\\
  0.0000& -5.0667&  5.8686& -1.9562&  0.0000&-14.9467&  5.8686& -1.9562\\
  6.8467&  2.9343&  2.6723& -2.5636&  6.8467&  2.9343&  2.6358& -2.4541\\
  5.8686& -1.9562&  0.0000& -5.0184&  5.8686& -1.9562&  0.0000& -4.7264
\end{array}\right) \ ,
\end{equation}
while that for the smeared operators is
\begin{equation}
\left(\begin{array}{rrrrrrrr} 
 10.2345& -5.1274& 12.5283&  5.3693& 10.7478& -6.6672& 12.5283&  5.3693\\
  0.0000& -5.1477& 10.7385& -3.5795&  0.0000& -9.2539& 10.7385& -3.5795\\
 12.5283&  5.3693& 10.3081& -5.3481& 12.5283&  5.3693& 10.2013& -5.0277\\
 10.7385& -3.5795&  0.0000& -5.7363& 10.7385& -3.5795&  0.0000& -4.8819
\end{array}\right) \ .
\end{equation}

For flavor $\xi_{345}$ pions, we need continuum operators
\begin{equation}
i = [\sum_\mu\sfno{\mu5}{345} \ \sfno{5\mu}{543}_{(I,II)},
     \sum_\mu\sfno{\mu}{345} \ \sfno{\mu}{543}_{(I,II)}] \ .
\end{equation}
and the lattice operators
\begin{eqnarray*}
j&=&[\sum_{\mu=3,4}\sfno{\mu5}{345}\ \sfno{5\mu}{543}_{(I,II)},
     \sum_{\mu=3,4}\sfno{\mu}{345}\ \sfno{\mu}{543}_{(I,II)},\\
&&\quad
     \sum_{\mu=1,2}\sfno{\mu5}{345}\ \sfno{5\mu}{543}_{(I,II)},
     \sum_{\mu=1,2}\sfno{\mu}{345}\ \sfno{\mu}{543}_{(I,II)}] \ .
\end{eqnarray*}
The matching matrices are
\begin{equation}
\left(\begin{array}{rrrrrrrr} 
  3.1770& -4.0775&  6.8467&  2.9343&  2.6615& -2.5312&  6.8467&  2.9343\\
  0.0000& -9.0555&  5.8686& -1.9562&  0.0000& -4.9320&  5.8686& -1.9562\\
  6.8467&  2.9343&  2.6615& -2.5312&  6.8467&  2.9343&  3.1770& -4.0775\\
  5.8686& -1.9562&  0.0000& -4.9320&  5.8686& -1.9562&  0.0000& -9.0555
\end{array}\right) \ ,
\end{equation}
\begin{equation}
\left(\begin{array}{rrrrrrrr} 
 10.4955& -5.9104& 12.5283&  5.3693& 10.2191& -5.0812& 12.5283&  5.3693\\
  0.0000& -7.2357& 10.7385& -3.5795&  0.0000& -5.0245& 10.7385& -3.5795\\
 12.5283&  5.3693& 10.2191& -5.0812& 12.5283&  5.3693& 10.4955& -5.9104\\
 10.7385& -3.5795&  0.0000& -5.0245& 10.7385& -3.5795&  0.0000& -7.2357
\end{array}\right) \ .
\end{equation}
for unsmeared and smeared operators respectively.

For scalar, pseudoscalar and tensor operators of flavor $\xi_{35}$
\begin{equation}
i= [\ixi{35}\ \ixi{53}_{(I,II)},
    \sfno5{35}\ \sfno5{53}_{(I,II)},
    \sum_{\mu<\nu}\sfno{\mu\nu}{35}\ \sfno{\nu\mu}{53}_{(I,II)}] \ ,
\end{equation}
matching requires the following operators
\begin{eqnarray*}
j&=&[\ixi{35}\ \ixi{53}_{(I,II)},
     \sfno5{35}\ \sfno5{53}_{(I,II)},\\
&&\quad  
     \sum_{\mu\ne3}\sfno{3\mu}{35}\ \sfno{\mu3}{53}_{(I,II)},
     \sum_{\mu\ne3}\sfno{3\mu5}{35}\ \sfno{5\mu3}{53}_{(I,II)}] \ .
\end{eqnarray*}
The results for unsmeared and smeared operators, respectively, are
\begin{equation}
\left(\begin{array}{rrrrrrrr} 
 -8.9254& 10.7553&  0.7500& -0.2500&  2.0878&  0.8948&  2.0878&  0.8948\\
  0.0000& 23.3405&  0.0000&  0.0000&  1.7895& -0.5965&  1.7895& -0.5965\\
  0.7500& -0.2500& -5.7497&  1.2283&  2.0878&  0.8948&  2.0878&  0.8948\\
  0.0000&  0.0000&  0.0000& -2.0648&  1.7895& -0.5965&  1.7895& -0.5965\\
 12.5267&  5.3686& 12.5267&  5.3686&  6.8395& -5.9070&  5.2835& -1.2389\\
 10.7371& -3.5790& 10.7371& -3.5790&  0.0000&-10.8816&  0.0000&  1.5667
\end{array}\right) \ ,
\end{equation}
\begin{equation}
\left(\begin{array}{rrrrrrrr} 
 -9.1536& 12.0827&  0.7500& -0.2500&  3.9817&  1.7064&  3.9817&  1.7064\\
  0.0000& 27.0947&  0.0000&  0.0000&  3.4128& -1.1376&  3.4128& -1.1376\\
  0.7500& -0.2500& -8.1815&  9.1665&  3.9817&  1.7064&  3.9817&  1.7064\\
  0.0000&  0.0000&  0.0000& 19.3181&  3.4128& -1.1376&  3.4128& -1.1376\\
 23.8899& 10.2385& 23.8899& 10.2385& 17.2133&-10.4124& 17.1090&-10.0995\\
 20.4771& -6.8257& 20.4771& -6.8257&  0.0000&-14.0240&  0.0000&-13.1896
\end{array}\right) \ .
\end{equation}

Finally, for operators with flavor $\xi_{345}$, it is convenient
to first present results for scalars and pseudoscalars, i.e.
\begin{equation}
i = [\ixi{345}\ \ixi{543}_{(I,II)},
     \sfno5{345}\ \sfno5{543}_{(I,II)}] \ ,
\end{equation}
which requires the lattice operators
\begin{equation}
j = [\ixi{345}\ \ixi{543}_{(I,II)},
     \sfno5{345}\ \sfno5{543}_{(I,II)}] \ ,
     \sum_{\mu<\nu}\sfno{\mu\nu}{345}\ \sfno{\nu\mu}{543}_{(I,II)}] \ .
\end{equation}
The results for unsmeared and smeared operators, respectively are
\begin{equation}
\left(\begin{array}{rrrrrr} 
 -8.1285&   8.3646&   0.7500&  -0.2500&   2.0878&   0.8948\\
  0.0000&  16.9653&   0.0000&   0.0000&   1.7895&  -0.5965\\
  0.7500&  -0.2500&  -8.1285&   8.3646&   2.0878&   0.8948\\
  0.0000&   0.0000&   0.0000&  16.9653&   1.7895&  -0.5965
\end{array}\right) \ ,
\end{equation}
\begin{equation}
\left(\begin{array}{rrrrrr} 
 -8.8985&  11.3174&   0.7500&  -0.2500&   3.9817&   1.7064\\
  0.0000&  25.0539&   0.0000&   0.0000&   3.4128&  -1.1376\\
  0.7500&  -0.2500&  -8.8985&  11.3174&   3.9817&   1.7064\\
  0.0000&   0.0000&   0.0000&  25.0539&   3.4128&  -1.1376
\end{array}\right) \ .
\end{equation}
For the continuum tensor operator
\begin{equation}
i= [\sum_{\mu<\nu}\sfno{\mu\nu}{345}\ \sfno{\nu\mu}{543}_{(I,II)}] \ ,
\end{equation}
the following lattice operators are required
\begin{eqnarray*}
j&=&[\ixi{345}\ \ixi{543}+\sfno5{345}\ \sfno5{543})_{(I,II)},
     \sum'_{\mu<\nu}\sfno{\mu\nu}{345}\ \sfno{\nu\mu}{543}_{(I,II)},\\
&&\quad
     \sfno{34}{345}\ \sfno{43}{543}_{(I,II)},
     \sfno{12}{345}\ \sfno{21}{543}_{(I,II)}] \ .
\end{eqnarray*}
Here $\sum'_{\mu<\nu}$ means the sum excluding the terms with
$\mu=1,\nu=2$ and $\mu=3,\nu=4$.
The results are
\begin{equation}
\left(\begin{array}{rrrrrrrr} 
 12.5267&  5.3686&  6.8395& -5.9070&  7.1507& -6.8404&  5.2835& -1.2389\\
 10.7371& -3.5790&  0.0000&-10.8816&  0.0000&-13.3707&  0.0000&  1.5667
\end{array}\right) \ ,
\end{equation}
\begin{equation}
\left(\begin{array}{rrrrrrrr} 
 23.8899& 10.2385& 17.1940&-10.3544& 17.2254&-10.4486& 16.8368& -9.2828\\
 20.4771& -6.8257&  0.0000&-13.8693&  0.0000&-14.1205&  0.0000&-11.0117
\end{array}\right) \ ,
\end{equation}
for smeared and unsmeared operators respectively.

In summary, we see that the corrections for operators with flavors
other than $\xi_5$ are of moderate size for all spins,
not exceeding 20\% for unsmeared operators, and 30\% for smeared
operators. 

\subsection{Converting results to NDR}
\label{sintondr}

To convert our results for $c_{ij}$ from \DREZpr\ to \NDR, we 
proceed in two steps. First, we convert to \NDRpr, using Eq.
\ref{schemeshifting}
\begin{equation}
c_{ij}(\NDRpr) = c_{ij}(\DREZpr) + d^T_{ij} - d^I_{ij} \ .
\end{equation}
Note that this conversion depends only on the spin, and not on the
flavor, of the operators.
In the second step we convert from \NDRpr\ to \NDR. 
We do this in the standard way
by comparing the one-loop matrix elements in the two schemes.
Since this is entirely a continuum calculation, it also
does not depend on the flavor of the operators.
We use the notation of Refs. \cite{newburas,martinelli}
\begin{equation}
c_{ij}(\NDR) = c_{ij}(\NDRpr) + \Delta r_{ij} \ .
\end{equation}

For the bilinears, $i,j=[(V-A)_I, (V-A)_{II}]$,
$i,j=[(V+A)_I, (V+A)_{II}]$, and
$i,j=[(S-P)_I, (S-P)_{II}]$, only the Xc diagrams contribute,
and the result is the same for all three cases
\begin{equation}
\Delta r_{ij} = 
\left(\begin{array}{rr}
  1 & -3 \\
 -3 &  1  
\end{array}\right) \ .
\end{equation}
This notation means that the result applies for any flavor, e.g.
$i,j= \left( \opoldc VSI, \opoldc VS{II}\right)$.
For $i,j=\left[(S+P)_I, (S+P)_{II}, T_I, T_{II} \right]$,
both Xb and Xc diagrams contribute and we find
\begin{equation}
\Delta r_{ij} = \frac16
\left(\begin{array}{rrrr}
  9  & -3 & 14 &  6 \\
  0  &  0 & 12 & -4 \\
-28  &-12 & -5 & -9 \\
-24  &  8 &-12 &  4
\end{array}\right) \ .
\end{equation}

This last result deserves some explanation.
Previous calculations using \NDR\ have considered only
four-fermion operators with spins $V\pm A$ and $S-P$.
To extend these calculations to $S+P$ and $T$
we must decide how to project against evanescent operators. 
In the notation of Ref. \cite{martinelli},
we use the projectors
\begin{equation}
P_1 = \frac1{32}\pl \otimes \pl \ , \quad
P_2 = \frac1{128} \sum_{\mu\nu} \g\mu \g\nu \pl \otimes \g\nu\g\mu \pl \ ,
\end{equation}
where the sums run over $n=4-2\epsilon$ dimensions.
Given a positive parity operator $\co$, 
whose composition in terms of $S+P$ and $T$ is unknown,
\begin{equation}
\co = a (S+P) + b (T) + {\rm evanescent\ operators} \ ,
\end{equation}
but whose projections are known, we use
\begin{equation}
\left(\begin{array}{r} a \\ b \end{array}\right) =
\frac{1}{36}
\left(\begin{array}{rr} 
  0 		& 18 + 18\epsilon \\
 24+ 28\epsilon & -6 - 13\epsilon	
\end{array}\right) 
\left(\begin{array}{r} P_1(\co) \\ P_2(\co) \end{array}\right) \ .
\end{equation}
We have dropped terms of $O(\epsilon^2)$.

We close this section with the example of $B_K$.
When calculated on the lattice using the pseudo-Goldstone kaons as
external states, this is the ratio of the matrix element of 
\begin{equation}
\co_B = [(V-A)\times P]_{I} + [(V-A)\times P]_{II} \ ,
\end{equation}
between a $K$ and a $\bar K$,
to the square of the matrix element of the bilinear 
$A_4=\sfno{45}5$ between a $K$ and the vacuum \cite{bkprl,book}.

Using the above, we find 
\begin{equation}
\co_B^\CONT(\NDR) =
\left[1 + {g^2\over 16\pi^2}(4\ln(\pi/a\mu)-\frac43)\right]\co_B^\LATT +
{g^2\over 16\pi^2} \delta\co \ ,
\end{equation}
where
\begin{equation}
\delta\co = 
  10.6702 \opoldc API + 1.6598 \opoldc AP{II} 
 -10.1426 \opoldc VPI - 7.4626 \opoldc VP{II} 
\end{equation}
for unsmeared operators, and
\begin{equation}
\delta\co =
  13.1912 \opoldc API + 10.3174 \opoldc AP{II}
 -13.0211 \opoldc VPI - 12.1878 \opoldc VP{II} 
\end{equation}
for smeared operators.
In these results, the effect of changing from \DREZpr\ to \NDR\ is
the coefficient $-4/3$ multiplying $\co_B^\LATT$. 
This conversion factor agrees with that of Ref. \cite{tsukubapert}.

For the denominator in $B_K$, the matching relation is the same in
\NDR\ and \DREZpr\ schemes
\begin{equation}
\label{bkdenomeqn}
(A_4^\CONT)^2 = (A_4^\LATT)^2 (1 + {g^2\over 16\pi^2} 2 C_F c_{VS})\ ,
\end{equation}
with $2 C_F c_{VS}= 0, -3.8769$ for unsmeared and smeared
operators respectively. 
Without tadpole improvement, each of the numerical coefficients in
both numerator and denominator is augmented by $-24.4661$.

We can combine the corrections to the numerator and denominator of $B_K$,
since the latter are an overall factor.
This amounts to subtracting the coefficient of $g^2$ in the denominator
from each of the numbers in the results for the numerator.
Doing this, we see that
the results with and without tadpole-improvement are the same.
This is because, for Landau-gauge operators, 
the tadpole diagrams are only on the external legs,
of which there are the same number for both numerator and denominator.

\newsection{PENGUIN DIAGRAMS}
\label{spenguin}

\subsection{Overview}

\begin{figure}
\centerline{\psfig{file=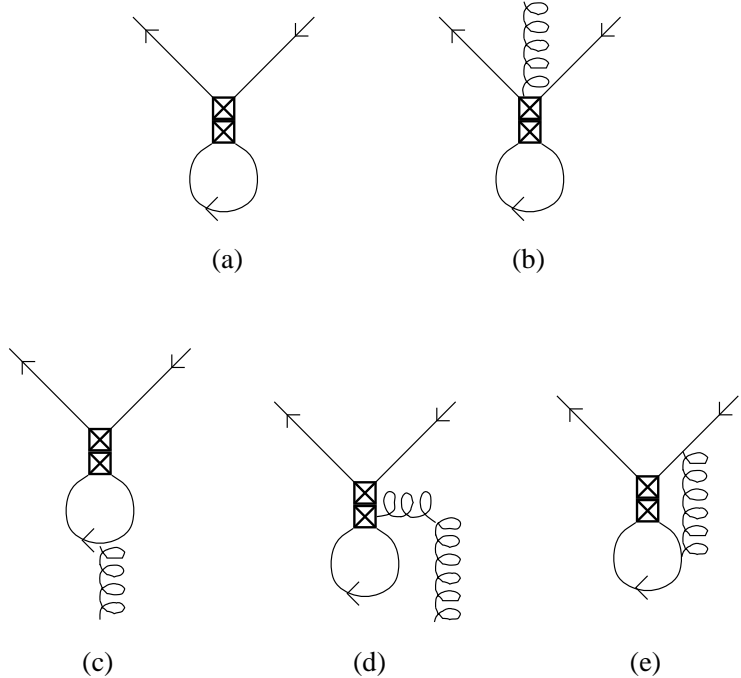,height=3.5truein}}
\caption[lowerdimfig]{\fc
Examples of diagrams leading to mixing of four-fermion operators
with bilinears. Each box represents a quark bilinear.
}
\label{lowerdimfig}
\end{figure}

We now turn to diagrams in which one of the quarks in the four-fermion 
operator is contracted with one of the antiquarks to form a closed loop.
Examples are shown in Fig. \ref{lowerdimfig}.
They give rise to mixing of four-fermion operators with quark bilinears 
(possibly containing gluon fields and derivatives).
The mixing coefficients are proportional to $a^{d'-d}$, 
where $d=6$ and $d'\ge3$ are the dimensions of the four-fermion operators
and bilinears, respectively. 
The appropriate treatment depends on $d'$:
\begin{itemize}
\item
For $d'<6$, i.e. mixing with lower dimension operators, the 
coefficients are proportional to inverse powers of the lattice spacing.
Such ultraviolet divergent coefficients
cannot be calculated to the required accuracy 
using perturbation theory \cite{nonpertref}.
One must, instead, use a non-perturbative method to subtract 
any lower dimension operators that are produced.
In the following section we provide a catalogue of
such operators, and show that only one of them contributes to
the matrix elements that are used in actual numerical calculations.
\item
For $d'=6$, a reliable perturbative calculation is possible.
We perform a one loop computation in this section.
\item
For $d'>6$, the calculation is sensitive to infrared physics
and thus unreliable. The mixing vanishes in the 
continuum limit, however, and can be ignored.
\end{itemize}

The one-loop diagrams are those shown in Figs. \ref{lowerdimfig}(a-d),
together with diagrams of the same form having
additional gluons attached to the bilinears.
These latter diagrams add no new features to the following analysis,
and we will not discuss them further.
Note that we have chosen to attach the quark loop to a single bilinear.
All continuum operators can be transcribed onto the lattice in this
way, and we use this method both because it is simpler, and 
because it has been used in all numerical simulations to date.
Given this choice, it turns out that the diagrams of Figs. 
\ref{lowerdimfig}(a) and (b) lead only to mixing with bilinears having $d'<6$.
The only one-loop diagrams giving $d'=6$ operators are those of Figs.
\ref{lowerdimfig}(c) and (d), and these we calculate in
this section. They also give rise to mixing with a $d'=5$ operator,
as discussed below.

\begin{figure}
\centerline{\psfig{file=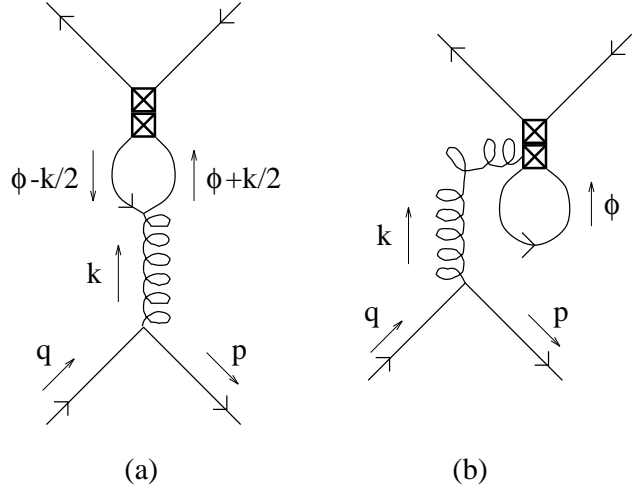,height=2.5truein}}
\caption[penguinfig]{\fc
``Penguin'' contributions to the renormalization of four-fermion operators.
Diagram (b) is absent for Landau gauge operators.
Each box represents a quark bilinear.
}
\label{penguinfig}
\end{figure}

In the continuum a considerable simplification occurs upon
using the equations of motion.
In the chiral limit, the $d'=6$ bilinears either vanish,
or can be converted into four-fermion operators.
This use of the equations of motion is equivalent to attaching a quark
line to the gluon, converting Figs. \ref{lowerdimfig}(c) and (d)
into the ``penguin'' diagrams shown in Fig. \ref{penguinfig}.
On the lattice, not all of the $d'=6$ bilinears that occur at one-loop can 
be rewritten as four-fermion operators using the equations of motion.
Nevertheless, it turns out that the operators of interest can be so
rewritten, so we couch our discussion in terms of a calculation
of lattice penguin diagrams. We first do the general calculation,
and then describe some applications in sect. \ref{spenapps}.

We concentrate on gauge-invariant operators in this and the following section.
Such operators receive corrections from both
Fig. \ref{penguinfig}(a),
in which a gluon connects the closed quark loop with a spectator quark line,
and from Fig. \ref{penguinfig}(b), 
in which the gluon connects directly to the bilinear.
The latter diagram is absent in the continuum;
on the lattice it serves to cancel mixing with lower dimension
gauge non-invariant operators induced by Fig. \ref{penguinfig}(a).
This cancellation does not occur for Landau-gauge operators,
for which Fig. \ref{penguinfig}(b) is absent.
This illustrates the problem with using Landau-gauge operators for
diagrams involving closed quark loops: they
mix with gauge non-invariant lower dimension bilinears.
This requires additional subtractions, and makes it difficult to extract 
phenomenologically interesting matrix elements.
We explain this further in sect. \ref{landaupen}.

At higher orders, diagrams of the type shown in Fig. \ref{lowerdimfig}(e)
lead to mixing with bilinears both of lower dimension and 
with a large number of additional operators having $d'=6$.
Although we have not calculated the mixing coefficients,
we discuss in sect. \ref{beyondoneloop}
the most interesting of the operators that arise.
Of course, there is also mixing with $d'=6$ operators due to X diagrams,
which we have considered above.

Finally, we note that the results of this section are 
unaffected by tadpole improvement, because 
one-loop tadpole diagrams are already accounted for when calculating
X-diagrams.

\subsection{Notation}

Penguin diagrams occur for operators which transform
as octets under the continuum flavor SU(3) symmetry
(or which contain a part which so transforms).
Aside from overall factors, which we discuss in sect. \ref{spenapps},
the calculation is the same for all octet operators, 
so we consider in detail only the operator
\begin{equation}
\co_{S'F',SF} = \chibar_s \sfno {S'}{F'}\chi_d \ 
                \chibar_u \sfno SF \chi_u
\ ,
\end{equation}
where color indices are suppressed.
The continuum flavors have been chosen so that there is only one possible
contraction leading to a penguin diagram, that having a $u$ quark in the loop.
Note that we consider the most general form of the four-fermion operator,
in which the spins and flavors of the two bilinears are different.
This requires no extra work for penguin diagrams because the first bilinear
is a spectator in the calculation. 

The continuum operator corresponding to $\co_{S'F',SF}$ is
\begin{equation}
\co_{S'F',SF}^{\CONT}= \bar S \opergx {S'}{F'} D \
		       \bar U \opergx  S   F   U \ .
\end{equation}
By construction, the tree level matrix elements of these two operators,
calculated respectively on the lattice and in the continuum,
are the same for physical momenta\footnote{%
As discussed in PS, this is true when the continuum matrix element
is expressed in the appropriate spin-flavor basis.}
\begin{equation}
\label{treeeq}
\vev{\co_{S'F',SF}^\CONT}^{(0)} = \vev{\co_{S'F',SF} }^{(0)} 
(1+ O(p_{\rm phys} a)) \ .
\end{equation}
The superscript refers to the number of loops.
Both lattice and continuum theories differ from QCD by having an additional
factor of $N_f=4$ flavors.
In particular, there are $N_f$ quarks running around the loop,
rather than a single quark in QCD.
To match onto QCD, one must divide the diagram by $N_f$. 
To make it possible to see the effect of this correction,
we keep the factor of $N_f$ explicit below.

One-loop diagrams lead to mixing with a single class of
SU(3) octet operators, which we call ``penguin operators''.
The continuum and lattice forms are, respectively,
\begin{eqnarray}
\nonumber
\cp_{S'F',S''F''} &=& 
\chibar_s \sfno {S'}{F'} \chi_d\ 
\sum_q  \chibar_q \sfno {S''}{F''} \chi_q \ , \\
\label{contpenopdef}
\cp_{S'F',S''F''}^\CONT &=& 
\bar S \opergx {S'}{F'} D\ \sum_Q  \bar Q \opergx {S''}{F''} Q \ .
\end{eqnarray}
The sums run over the number of active light flavors,
typically $u,d,s$ and $c$ in present lattice simulations.

The color factors are simple to calculate.
All the operators come in two color types, and the matching matrix
is a $2\times 2$ matrix in color space.
Using the notation of sect. \ref{scolorfactors}, we find
\begin{equation}
\label{pencorreqn}
\bfco^\CONT_i = \bfco^\LATT_i 
+ {g^2\over 16\pi^2} \sum_j 
  \rho_{ij} \stackrel{\longleftrightarrow}{C}_{\rm\!\! pen}
   \bfcp_j + {\rm X\ diagram\ contributions} \ ,
\end{equation}
where $i$ and $j$ represent both the spin and flavor indices of the operators,
and 
\begin{equation}
\label{pencolor}
\stackrel{\longleftrightarrow}{C}_{\rm\!\! pen}
	  = \frac16\left( \begin{array}{rr} 
                   3 & -1 \\ 0 & 0 \end{array} 
            \right)
\end{equation}
is the color matrix.
The first row follows from Eq. \ref{colorfierz}, while
the second row vanishes because the gauge generators are traceless.

To calculate $\rho$ we need the lattice and continuum one-loop matrix
elements. These have the general form
(in a notation based on that of Ref. \cite{uclapert})
\begin{equation}
\label{lattpeneq}
\vev{\co_{S'F',SF} }_\PEN^{(1)}
= {g^2\over 16\pi^2} \sum_{S''F''}
\rho_{SF,S''F''}^\LATT \vev{ \cp_{S'F',S''F''} }^{(0)} 
\end{equation}
(the subscript $\PEN$ indicating that only the penguin part of the
matrix element is included), and
\begin{equation}
\label{contpeneq}
\vev{\co_{S'F',SF}^{\CONT} }_\PEN^{(1)}
= {g^2\over 16\pi^2} \sum_{S''F''} \rho_{SF,S''F''}^\CONT
\vev{ \cp_{S'F',S''F''}^\CONT }^{(0)} \ ,
\end{equation}
respectively.
This is a rather elaborate notation for the continuum result,
because, when $ma=0$, almost all of the one-loop coefficients vanish
\begin{equation}
\rho_{SF,S''F''}^\CONT = \sum_\mu \delta_{S\vphantom{'} \widehat\mu}\ 
\delta_{F\vphantom{'} I}\ \delta_{S' \widehat\mu}\ \delta_{F' I}\ 
\rho_{\widehat\mu I,\widehat\mu I}^\CONT \ .
\end{equation}
Such a notation is required, however, for the lattice matrix elements.
Combining Eqs. \ref{lattpeneq} and \ref{contpeneq}, 
and using Eq. \ref{treeeq}, we find 
\begin{equation}
\rho_{ij} = \rho_{ij}^\CONT - \rho_{ij}^\LATT \ .
\end{equation}
This notation is slightly redundant in that $\rho_{ij}$ depend only
on $S$, $F$, $S''$ and $F''$, but not on $S'$ and $F'$.

\subsection{Calculation}

Since the first bilinear in Fig. \ref{penguinfig} is simply a spectator,
we are, in effect, calculating the mixing of the second bilinear,
$\sfno SF$, into other bilinears.
We first consider this mixing on the lattice.
Using the Feynman rules given in PS, 
and the momentum definitions from Fig. \ref{penguinfig}, we find \footnote{%
In this subsection the sum over quark flavors $q$ (see Eq. \ref{contpenopdef})
for the lower quark line in Fig. \ref{penguinfig} is implicit.}
\begin{equation}
\label{penbilineq}
  \vev{ \sfno SF }_\PEN^{(1)} = {ig \over 16\pi^2} 
  \sum_{AB} \frac{1}{16} \sf S F A B  e^{{-i} k \cdot (A+B)/2}
  \sum_{\mu\nu} P_\mu(k;B,A) D_{\mu\nu}(k) V_\nu(k) \ .
\end{equation}
Here and in the following we have suppressed all color matrices and indices,
since the color factors have already been calculated.
The factors $P$, $D$ and $V$ represent respectively the penguin loop,
the gluon propagator and the lower quark-antiquark-gluon vertex.
Expressions for $D$ and $V$ are given below.
We first focus on the penguin loop, for which we find
\begin{eqnarray}
\label{pamp}
  P_\mu(k;B,A) &=& {N_f\over4}
\int_\phi (I_1 - I_2) e^{i \phi \cdot (A-B) } \ , \\
  I_1 &=& \sum_{CD} S(\phi+\half{k})_{BC}\ c_\mu \gam\mu_{CD}
                    S(\phi-\half{k})_{DA} \\
  I_2 &=& \sum_{CD} S(\phi)_{BC}\ c_\mu \gam\mu_{CD} S(\phi)_{DA} \ ,
\end{eqnarray}
where $S$ is the quark propagator
\begin{equation}
  S(\phi)_{AB} = F(\phi)\ \left[m\iden + i\sum_\mu s_\mu \gam\mu \right]_{AB}
 \ ,
\quad  F(\phi) = \left[m^2 + \sum_\mu s_\mu^2\right]^{-1} \ ,
\end{equation}
and we are using the notations
\begin{equation}
  \int_\phi \equiv 16\pi^2 \int_{-\pi}^{\pi} {d^4\phi \over (2\pi)^4} \ ,
\quad  s_\mu \equiv \sin(\phi_\mu) \ ,
\quad  c_\mu \equiv \cos(\phi_\mu) \ .
\end{equation}

Various features of these results require explanation.
\begin{itemize}
\item
The factor of $1/16$ multiplying $\sfno S F$
in Eq. \ref{penbilineq} arises from the
fact that we integrate the momenta over the entire Brillouin zone {\em and}
trace over the spin-flavor indices. To avoid ``double'' counting
we have to divide the result by $16$.
\item
Infrared regularization is provided by the quark mass, which we
can only set to zero after the lattice and continuum results are combined.
\item
We have transformed the matrices to the single-bar representation,
which corresponds to physical position on a hypercube. This is the
most convenient prescription since it allows one to easily determine
the constraints on $A$ and $B$. We have used these constraints to
simplify Eq. \ref{pamp}.   
\item
The $I_2$ term in Eq. \ref{pamp} is the contribution of 
Fig. \ref{penguinfig}(b). We have written it so as to show explicitly
that $P_\mu$ vanishes when $k=0$.
\item
This vanishing is one of the consequences of the Ward identity,
\begin{equation}
\label{ginvWI}
  \sum_\mu \sin(\half k_\mu) P_\mu(k;B,A) = 0 \ ,
\end{equation}
which is satisfied by our expression for all values of $k$.
This identity follows from gauge invariance,
and is necessary for the gluon to remain massless.
\end{itemize}

It turns out that contributions to operator mixing come not only from
$k\sim0$, as in the continuum, but also from $k \sim C\pi$, where
$C$ is a non-zero hypercube vector. 
This is because a momentum close to $\pi$ is physical for staggered fermions.
We first consider $k\sim0$. 
The integrand $I_1$ has to be expanded out to $O(k^2)$. 
Tedious algebra leads to the result
\begin{eqnarray}
\nonumber
  P_\mu(k;B,A) &=& 
  i \epsilon_{\mu\rho\sigma\nu} k_\rho \sf{\sigma\nu}{\sigma 5} B A I_a 
- i \half m k_\rho\left[\gam{\mu\rho}_{BA}-\gam{\rho\mu}_{BA}\right] I_b \\
\label{pampzero}
&& \quad  + (k_\mu k_\nu -\delta_{\mu\nu} k^2) \gam\nu_{BA} I_c 
+ O(k^3) \ ,
\end{eqnarray}
where the indices $\rho$, $\sigma$ and $\nu$ are summed.
The integrals are
\begin{eqnarray}
\nonumber
  I_a &=& \frac{N_f}{4} \int F^2 c_1^2 c_2^2 s_3^2 
	= \frac{N_f}{4} \times 11.2291 \ , \quad
  I_b = \frac{N_f}{4} \int F^2 c_1^2 c_2^2 = 
	\frac{N_f}{4} \left(\int F^2 -40.7767 \right)\ ,  \\
 I_c &=& \frac{N_f}{4} \int \sixth F^2 \left[ 2 - 2 s_1^2 - s_1^2s_2^2 \right] 
= \frac{N_f}{4}\left( \frac13 \int F^2-9.5146 \right)\ .
\end{eqnarray}
Note that $I_a$ is finite,
while $I_b$ and $I_c$ diverge logarithmically as $m\to0$.

The first two terms in Eq. \ref{pampzero} correspond to mixing of
the bilinear with gluonic operators which cannot be rewritten
using the equations of motion. In other words, for these terms we
are calculating Figs. \ref{lowerdimfig}(c) and (d) rather than penguin
diagrams. For these terms we should drop
the factor of $D\times V$ in Eq. \ref{penbilineq}, and multiply by
the gluon polarization tensor $\epsilon_\mu$.

To analyze the $I_a$ term we must expand the
factor $e^{{-i} k \cdot (A+B)/2}$ in Eq. \ref{penbilineq}
up to linear order in $k$ around the center of the hypercube.
The leading term is unity, and gives rise to mixing between the bilinear
$\chibar\sfno{\sigma\nu}{\sigma5}\chi$ 
and the lower dimension gluonic operator $\widetilde{F}_{\sigma\nu}$
(the dual of the gluon field strength).
Reinstating the spectator bilinear, this is an example of mixing
with a $d'=5$ operator.
It is consistent with the lattice symmetries, but,
as explained in sect. \ref{smixing}, it does not contribute to
the matrix elements of practical interest.
For the $O(k)$ term in the expansion, one finds after some
algebra, that it leads to mixing between
$\chibar\sfno{\sigma5}{\sigma\nu}\chi$ 
and $D_\nu\widetilde{F}_{\sigma\nu}$.
Including the spectator bilinear, this corresponds to mixing with
a $d'=6$ operator. At one-loop,
this mixing does not affect the operators needed to study the 
$\Delta I=1/2$ rule (which are listed below in sect. \ref{spenapps}),
since none contain bilinears having flavor $\xi_F=\xi_{\sigma\nu}$.
It will affect these operators at two-loop order.
However, as explained in sect. \ref{smixing},
its contribution in matrix elements will be suppressed
by powers of $a$ due to its flavor.

The $I_b$ term corresponds to mixing of the
operator $\chibar \gam {\mu\rho} \chi$ ($\mu\ne\rho$) with
$m F_{\mu\rho}$, both of which have dimension 3. 
Such mixing also occurs in the continuum, 
and when one combines the lattice and continuum corrections
the infrared divergence in $I_b$ cancels with that in the corresponding 
continuum integral. Since the operator vanishes in the chiral limit,
it is dropped in renormalization group analyses \cite{buras},
and we follow this practice here.
In fact, for the operators needed to study the $\Delta I=1/2$ rule, 
the initial operator contains no terms with tensor bilinears,
so mixing with $m F_{\mu\rho}$ begins only at two loop order.
Thus its contribution is suppressed both by the quark mass and by
an additional power of $g^2/16\pi^2$.
The factor of $m$ appears because of the chiral symmetry of 
staggered fermions, and is not present for Wilson fermions \cite{sigmawilson}.

We are thus interested only in the term proportional to $I_c$,
which corresponds to the mixing of $\gam \nu$ with $D_\mu F_{\mu\nu}$,
and which can be rewritten using the equations of motion.
To extract the infrared divergence we use
\begin{equation}
  \int F^2 = 16 ( -\ln(4m^2a^2) - \gamma_E + F_{0000} ) + O((ma)^2) \ .
\end{equation}
Numerical values of the constants are given, for example, in PS.
%
%
We calculate the remaining finite piece of $I_c$ numerically, yielding,
in total
\begin{equation}
I_c = - {8 N_f \over 3}\ln(m a/\pi) - 8.8945 + O((ma)^2) \ .
\end{equation}
The error on the numerical constant is roughly $0.0001$.

To complete the amplitude, the gluon must be joined onto the
external fermion line. The result in Feynman and Landau gauges is the
same because of the Ward identity Eq. \ref{ginvWI}, so we use the
Feynman gauge propagator
\begin{equation}
  D_{\mu\nu}(k) = {\delta_{\mu\nu} \over \sum_\rho 4 \sin^2(k_\rho/2)}
 \ .
\end{equation}
For $k\sim0$ the vertex is (see PS)
\begin{equation}
\label{lowervertexeq}
  V_\mu(p) = -ig \bar\delta(p' + k - q') \ggam\mu_{A'B'} 
  \cos(q'_\mu-\half k_\mu)
\ .
\end{equation}
The external momenta have been decomposed into physical 
and flavor-related parts:
$p = p' + A'\pi$ and $q=q' + B'\pi$,
where $p',q' \in [-\pi/2,\pi/2]$.
We assume that $p'$ and $q'=k+p'$ are both close to zero,
so that the cosine can be set equal to unity.
Using the equations of motion on the vertex 
(i.e. $\sum_\mu \sin(k_\mu/2) V_\mu = 0$ for massless fermions), we see that
only the $k^2$ term in the coefficient of $I_c$ in Eq. \ref{pampzero} survives.
This cancels the denominator of the gluon propagator, so that there is
no longer any dependence on $k$, and we have arrived at a local operator.
We can now do the integral over $k$ to remove the delta function in
Eq. \ref{lowervertexeq}, and sum over $A$ and $B$ in Eq. \ref{penbilineq}.
We can set the factor $\exp(-ik\cdot(A+B)/2)$ to unity, since the corrections
lead to operators with $d'>6$.
Thus the only non-zero matrix element from gluon momenta $k\approx 0$ is
\begin{equation}
\label{nr0mixing}
 \vev{ \gam \mu }_\PEN^{(1)} = - {g^2 \over 16\pi^2} I_c \ggam\mu_{A'B'} \ .
\end{equation}
Since $\ggam\mu_{A'B'}$ is the tree level matrix element of the
bilinear $\gam\mu$, we can read off one contribution to $\rho^\LATT$
\begin{equation}
\rho_{\widehat\mu I,\widehat\mu I}^\LATT = - I_c \ .
\end{equation}
Note that this contribution is diagonal in spin and flavor.

The corresponding continuum calculation is straightforward
(see Ref. \cite{uclapert} for more details), with the result
\begin{eqnarray}
 \vev{ \co_{SF}^\CONT }_\PEN^{(1)} &=& 
   {ig \over 16\pi^2} \sum_{\mu\nu\rho}
    (k_\mu k_\rho - \delta_{\mu\rho} k^2) \ D_{\mu\nu}(k) V_\nu(k) \ 
    {\Tr(\gamma_S \gamma_\rho)\over 4} {\Tr(\xi_F)\over N_f} \ I_c^\CONT \\ 
&=& - {g^2 \over 16\pi^2} I_c^\CONT 
\sum_\mu \delta_{S,\widehat\mu}\delta_{F,I} \opergi\mu \ .
\end{eqnarray}
The continuum integral is
\begin{equation}
  I_c^\CONT = - {4 N_f \over3} \ln({m^2\over\mu^2}) \ ,
\end{equation}
%
in \DREZ, \DRED, \NDR, \HV\ and \PV\ schemes\footnote{%
In the PV scheme $\mu$ is the fermion regulator mass.}.
Thus the only non-zero elements of the continuum mixing matrix are
\begin{equation}
\rho_{\widehat\mu I,\widehat\mu I}^\CONT = - I_c^\CONT \ .
\end{equation}
The infrared logarithms cancel when we combine lattice and continuum results
\begin{equation}
\label{finalrho}
\rho_{\widehat\mu I,\widehat\mu I} = I_c - I_c^\CONT
= - \frac{8 N_f}{3} \ln(\mu a/\pi) - 8.8945 \ .
\end{equation}
The additional contributions discussed below do not change this mixing
coefficient.

\subsection{Contributions from high momentum gluons}

As mentioned earlier, this is not the whole story. 
Unlike in the continuum, gluons with  momenta $k\approx C\pi$ 
($C\ne0$) also contribute on the lattice. 
This is because, while $p'$ and $q'$ must be close to zero,
$k=q'-p' + \pi(B'-A')$ need not be.
These contributions result in a local operator because both the gluon
propagator and the fermion loop are both short ranged.

It is instructive to begin by considering the vertex connecting
the gluon to the external line 
\begin{equation}
\label{vertexcpi0}
  V_\nu(C\pi) = -ig \bar\delta(p' - q') {(1 + (-1)^{C_\nu}) \over 2}
                \ggam\nu_{A',B'+C} \ .
\end{equation}
Clearly we must choose $C_\nu=0$ for the vertex to be non-vanishing.
Notice the appearance of the subscript $B'+C$ in place of $B'$.
It is simple to show that       
\begin{equation}
\label{vertexcpi}
  V_\nu(C\pi) = -ig \bar\delta(p' - q') {(1 + (-1)^{C_\nu}) \over 2}
                \ssf{\nu \tC}{\tC}{A'}{B'} \ ,
\end{equation}
where $\tC_\nu =_2 \sum_{\rho\ne\nu} C_\rho$.
This shows that the external quarks ``see'' this vertex as having 
a different spin and flavor.

The expressions \ref{vertexcpi0} and \ref{vertexcpi}
differ from the simplified Feynman rules listed in PS.
The rules given there are
appropriate for propagators and vertices involved in loop integrations,
but are inapplicable if more than one of the legs of the vertex is an
external line.
This is because, to derive the rules, one has to use the freedom to
redefine loop variables. For tree level couplings
no rearrangement is allowed and the correct vertex is as given above.
Note that similar changes occur for matrix elements of operators
when they are inserted at non-zero momenta.

The penguin loop is straightforward to calculate using the      
fact that the integration variable can be shifted freely.
Furthermore, since the Ward Identity Eq. \ref{ginvWI} holds for all $k$,
the longitudinal part of the gluon propagator does not contribute,
so that $D_{\mu\nu}\propto \delta_{\mu\nu}$. Thus in evaluating $P_\mu$
we can assume that $C_\mu=0$.
Including the momentum factor from the bilinear vertex,
the loop becomes
\begin{eqnarray}
\nonumber
   e^{-i k\cdot(A+B)/2} \ P_\mu(C\pi;B,A)
  &=& -m I_d \sum_{\rho\ne\mu}
    (1 - (-1)^{C_\rho}) \sf{\mu5\tC}{\rho5\tC} B A  \\
\nonumber
  & & \quad + \half I_e \sum_{\rho\ne\mu,\sigma\ne\mu} 
    ((-1)^{C_\rho} - (-1)^{C_\sigma})     
    \sf{\mu\tC}{\sigma\rho\tC} B A \\
\label{penloopCpi}
  & &\quad + I_d \sf{\mu\tC}{\tC} B A      
    \sum_{\rho\ne\mu} ((-1)^{C_\rho} - 1) \ .
\end{eqnarray}
The integrals needed are
\begin{equation}
  I_d = \frac{N_f}{4}\int F^2 c_1^2 s_2^2 =\frac{N_f}{4}\times16.3107\ , \quad
  I_e = \frac{N_f}{4}\int F^2 c_1^2 s_2^2 s_3^2 = \frac{N_f}{4}\times5.0816 \ ,
\end{equation}
both of which are infrared finite.
The first term in Eq. \ref{penloopCpi}
corresponds to mixing with a $d'=7$ operator,
because of the factor of $m$, and can be dropped.
The remaining two terms correspond to mixing with $d'=6$ operators.

To complete the calculation we need the gluon propagator
\begin{equation}
  B^{-1} (C\pi) = \sum_\rho 2(1 - (-1)^{C_\rho}) \ .
\end{equation}
Altogether, after integrating over gluon momenta near
$k= C\pi$ to remove the delta-function in Eq. \ref{vertexcpi}, 
we find the following non-zero matrix elements
\begin{eqnarray}
\label{diagCpi}
  \vev{ \sfno{\mu\tC}{\tC} }_\PEN^{(1)} &=& - {g^2\over 16\pi^2} 
	\half I_d \ssf{\mu\tC}{\tC}{A'}{B'} \ , \\
\label{offdiagCpi}
  \vev{ \sfno{\mu\tC}{\sigma\rho\tC} }_\PEN^{(1)} &=&  {g^2\over 16\pi^2} 
  \half I_e B(C\pi) \left[ (-1)^{C_\rho} - (-1)^{C_\sigma} \right]     
	\ssf{\mu\tC}{\tC}{A'}{B'} \ ,
\end{eqnarray}
where $\mu$, $\rho$ and $\sigma$ can take any values, but are all different,
and $C$ is any non-zero hypercube vector satisfying $C_\mu=0$.
The remaining one-loop contributions to $\rho_{ij}^\LATT$ 
can be read off from this equation.
There are no continuum contributions to
these coefficients, so $\rho_{ij} = - \rho_{ij}^\LATT$.

These results show the complications due to the use of staggered fermions.
The mixing is between bilinears all of which have $F \ne I$; the flavor
is transmitted by the highly off-shell gluon.
Furthermore, the mixing of Eq. \ref{offdiagCpi} is flavor off-diagonal,
but is nevertheless consistent with the hypercube symmetry group 
\cite{verstegen}, as we have checked explicitly.

\subsection{Applications}
\label{spenapps}

The continuum octet operators which are needed to calculate kaon decay
amplitudes are 
\begin{eqnarray}
Q_1 &=& \sbar_a \g\mu \pl u_b\ \ubar_b \g\mu\pl d_a \ ,  \nonumber\\
Q_2 &=& \sbar_a \g\mu \pl u_a\ \ubar_b \g\mu\pl d_b \ ,  \nonumber\\
Q_3 &=& \sbar_a \g\mu \pl d_a\ \sum_{q}\ \qbar_b \g\mu\pl q_b \ , \nonumber\\
\label{burasops}
Q_4 &=& \sbar_a \g\mu \pl d_b\ \sum_{q}\ \qbar_b \g\mu\pl q_a \ , \\
Q_5 &=& \sbar_a \g\mu \pl d_a\ \sum_{q}\ \qbar_b \g\mu\pr q_b \ , \nonumber\\
Q_6 &=& \sbar_a \g\mu \pl d_b\ \sum_{q}\ \qbar_b \g\mu\pr q_a \ . \nonumber
\end{eqnarray}
where we follow the notation of Refs. \cite{newburas,martinelli},
except that we use $\pl$ for the the left-handed projector.
The sum over $q$ runs over the $f$ active flavors: 
$u$, $d$, $s$ and possibly $c$.
For brevity we have left out the electromagnetic penguin operators,
$Q_{7-10}$. It is straightforward to extend the discussion to them.

References \cite{newburas,martinelli} have calculated the two-loop anomalous
dimension matrices for these operators in the \NDR\ scheme.
(They also use the \HV\ scheme which we do not consider here.)
We wish to find the lattice operators 
which match onto $Q_{1-6}$ to one-loop accuracy.
We consider here only the contributions to the matching from penguin diagrams;
those from X diagrams can be obtained from sect. \ref{sxresults} or, 
in the case of gauge-invariant operators, from Ref. \cite{tsukubapert}.

The first step is entirely in the continuum.
It turns out to be more convenient to change from the basis of operators
given above to that in which $Q_1$, $Q_2$, $Q_3$ and $Q_5$ are replaced by
their Fierz transforms
\begin{eqnarray}
\tQ_1 &=& \sbar_a \g\mu \pl d_a\ \ubar_b \g\mu\pl u_b \ ,  \nonumber\\
\tQ_2 &=& \sbar_a \g\mu \pl d_b\ \ubar_b \g\mu\pl u_a \ ,  \nonumber\\
\label{newbasis}
\tQ_3 &=&  \sum_{q}\ \sbar_a \g\mu \pl q_b\ \qbar_b \g\mu\pl d_a \ , \\
\tQ_5 &=& -2 \sum_{q}\ \sbar_a \pr q_b\ \qbar_b \pl d_a \ , \nonumber
\end{eqnarray}
respectively. The two bases are identical in four dimensions,
but differ in n-dimensions. In particular, the finite parts of their
penguin matrix elements differ. 
The matching coefficients between these two bases are simple to obtain
by generalizing the work of Ref. \cite{newburas}, 
and we give them below.

When we transcribe these operators onto the lattice, 
we do so at the level of contractions \cite{wius,book}. 
For example, the penguin contraction of $\tQ_2$ in a $K\to\pi\pi$ matrix
element is obtained using the penguin contraction of the lattice operator
\begin{equation}
\label{tq2latt}
\tQ_2^\LATT = 
\frac{1}{N_f} \left[
(\chibar_s \sfno {\mu5}{5}\chi_d \ \chibar_u \gam\mu \chi_u)_{I} +
(\chibar_s \sfno {\mu}{5}\chi_d \ \chibar_u \gam{\mu5} \chi_u)_{I}
\right] \ .
\end{equation}
The $1/N_f$ cancels the extra lattice flavors in the loop.
In all cases of practical interest
the flavors of the bilinears are either $I$ or $\xi_5$ \cite{epsilonpus}.
This means that the mixings of Eqs. \ref{diagCpi} and \ref{offdiagCpi}
are not relevant, because the flavor of the bilinears in these equations
is never $I$ or $\xi_5$. The only relevant mixing is that of 
Eq. \ref{nr0mixing}, which affects just the bilinears $\gam\mu$.
Thus only the first term in $\tQ_2^\LATT$ receives corrections.

We can now explain why we use $\tQ_2$ rather than $Q_2$. 
When we transcribe the former onto the lattice, as in Eq. \ref{tq2latt},
the penguin loop attaches to a single bilinear,
and the spin-flavor indices are contracted in two loops.
On the other hand, had we transcribed $Q_2$ onto the lattice, 
the spin-flavor indices would have been contracted into a single loop.
While there is no fundamental reason to prefer one choice over the other,
numerical simulations have so far used the two spin-flavor loop form
\cite{epsilonpus}. Thus we have calculated the corrections in sect.
\ref{spenguin} for this form.
Given this, we must match onto $\tQ_2$ rather than $Q_2$.
A similar argument applies for $Q_1$, except that
the penguin contraction vanishes at one-loop because of the color factor, 
so it turns out to make no difference whether we choose $Q_1$ or $\tQ_1$.

For the operators $Q_{3-6}$, the situation is more complicated.
Each operator has two types of penguin contraction:
either $\bar q$ is contracted with $q$, or
the penguin loop is formed by contracting $q$ with $\sbar$ or $\qbar$ with $d$.
For each operator, one type of contraction has a vanishing color factor.%
\footnote{%
We stress that in an actual numerical simulation one must include both
types of contractions---only at one-loop order does one of them vanishes.
}
To put the remaining contraction in two spin-flavor loop form, 
we must match onto $\tQ_3$, $Q_4$, $\tQ_5$ and $Q_6$.
The lattice transcriptions of the penguin contractions of these operators
are analogous to Eq. \ref{tq2latt}, and we do not give explicit forms.

We now collect the penguin contributions to the matching coefficients.
For the negative parity parts of the operators, the general form is
\begin{eqnarray}
\label{contopmatch}
Q_i(\NDR) &=& Q_i^\LATT + P[Q_i] {g^2\over 16\pi^2} Q_P^\LATT \ , \\
Q_P^\LATT &=& \frac{2}{N_f} \sum_{q} \left[
(\chibar_s \sfno {\mu5}{5}\chi_d \ \chibar_q \gam\mu \chi_q)_{I} -
\frac13(\chibar_s \sfno {\mu5}{5}\chi_d \ \chibar_q \gam\mu \chi_q)_{II}
\right] \nonumber \ .
\end{eqnarray}
For the positive parity parts of the operators, the same result applies
with $\sfno{\mu5}{5}$ replaced with $\gam\mu$ in $Q_P^\LATT$.
The calculation done in the previous two subsections applies directly to
the operator $\tQ_{2}$. For the other operators we need only count the 
number of continuum flavors that can run round the penguin loop.
Including the matching coefficient between the continuum bases, we find
\begin{equation}
P[Q_1]=0 \ ,\ 
P[Q_2]=\frac13 + \frac14\rho \ ,\ 
P[Q_3]=2 P[Q_2] \ ,\ 
P[Q_4]= P[Q_6] = \frac14 f \rho \ ,\ 
P[Q_5]=0 \ ,\ 
\end{equation}
where $\rho\equiv\rho_{\widehat\mu I,\widehat\mu I}$.
One check on Eqs. \ref{finalrho} and \ref{contopmatch} is that we can
extract the contribution of penguin diagrams to the 
one-loop anomalous dimension matrix from the coefficient
of $\ln(\mu)$. Our results agree with those
of, for example, Ref. \cite{newburas}.

How large are the matching corrections?
First we note that $Q_P=-Q_3/3+Q_4-Q_5/3+Q_5$. Thus the
operator produced by mixing is of the same ``size'' as the original
operators. To estimate the size of the coefficients,
we proceed as above and set $\mu =\pi/a$ and $g^2/16\pi^2\approx 1/90$. 
The largest coefficient of $Q_P$ is
$(g^2/16\pi^2) f \rho/ (2 N_f) \approx -1/20$.
Thus the penguin contributions to matching are small.

There is a subtlety in the application of Eq. \ref{contopmatch}
to the quenched approximation. Recall that the initial mixing is with
a bilinear containing gluon fields, which is then rewritten as a
four-fermion operator using the equations of motion.
The second step is altered in the quenched approximation.
It is simple to show that,
when taking matrix elements of the operator produced by mixing
one should exclude contractions between the $\chibar_q$ and $\chi_q$ fields.

\subsection{Beyond one loop}
\label{beyondoneloop}

At higher orders in perturbation theory the continuum operators
$Q_{1-6}$ will match onto a linear combination of a much larger number
of $d'=6$ lattice operators. 
We wish to point out some general features of this mixing.
Some of the results stated here rely on the analysis of sect. \ref{smixing}.

Mixing produces both four-fermion and bilinear $d'=6$ operators.
We are only interested in those operators whose flavors match those
of the external states, i.e. those having flavor $\xi_F=1$ or $\g5$. 
One-loop mixing produces essentially all the four-fermion operators 
satisfying this criterion: higher loops hold no surprises.
This is not true, however, for the bilinears.
In particular, the lattice transcriptions of the operators
\begin{equation}
\label{noncovops}
\sum_\mu \sbar \g\mu\pl {\stackrel{\longrightarrow}{D_\mu}} 
{\stackrel{\longrightarrow}{D_\mu}} {\stackrel{\longrightarrow}{D_\mu}} 
d \quad 
{\rm and}\quad
\sum_\mu \sbar {\stackrel{\longleftarrow}{D_\mu}}
{\stackrel{\longleftarrow}{D_\mu}}
{\stackrel{\longleftarrow}{D_\mu}}
\g\mu \pl d \ ,
\end{equation}
can appear. These are allowed by the lattice symmetries, though
not by the continuum symmetries, and cannot be simplified using the
equations of motion. Normally such operators are suppressed 
by powers of $a$, but this is not true here because they are corrections
to $d'=4$ operators which themselves are multiplied by $a^{-2}$.
The operators of Eq. \ref{noncovops} are not produced by one loop graphs 
of Figs. \ref{lowerdimfig} and \ref{penguinfig}, because these do not depend on
the momenta of the $s$ and $d$ quarks.
Presumably they appear in two loop graphs such as that shown in
Fig. \ref{lowerdimfig}(e).

Because they appear only in two-loop graphs, these operators are likely
to be unimportant numerically. We mention them mainly as a curiosity.
We note, however, that they also occur with Wilson fermions,
yet have so far been overlooked, while other operators appearing at
two-loops have been considered \cite{sigmawilson}.
The difference of the two operators in Eq. \ref{noncovops} is even
under CPS symmetry (see Ref. \cite{cps} and the sect. \ref{smixing}),
and so does contribute to both even and odd parity matrix elements.
Thus the tricks suggested in Ref. \cite{cpskpipi} do not remove this 
operator.

We mention also two other $d'=6$ operators which appear at two (or higher)
loops
\begin{equation}
\label{totderivs}
\sum_{\mu,\nu} \partial_\mu\left[ \sbar \g\nu F_{\mu\nu} \pl d \right]
 \quad {\rm and}\quad
\sum_{\mu,\nu} 
\partial_\mu\left[ \sbar \g\nu\g5 {\widetilde F}_{\mu\nu} \pl d \right]\ .
\end{equation}
These are allowed by the continuum symmetries, and by CPS, 
but are usually dropped because, being total derivatives, they do not
contribute to physical matrix elements.
On the lattice, however, the standard approaches to calculating
$K\to\pi\pi$ amplitudes use indirect methods based on amplitudes
in which the weak Hamiltonian inserts momenta.
For example, one method requires the $K\to\pi\pi$ amplitude at
threshold for equal mass quarks, so that energy is inserted by the operator
\cite{cpskpipi}.
Thus the operators of Eq. \ref{totderivs} can contribute.

\subsection{Landau gauge operators}
\label{landaupen}

We close this section with a brief discussion of the differences
in the results for Landau-gauge and gauge-invariant operators.
The only difference in the one-loop calculation is that
Fig. \ref{penguinfig}(b) is absent for Landau gauge operators,
leading to a violation of the Ward identity, Eq. \ref{ginvWI}.
The contribution of this diagram to $P_\mu(k;B,A)$ is (for $ma=0$)
\begin{equation}
\label{diagramtwo}
  \int_\phi S(\phi)_{BA} (A\!-\!B)_\mu e^{i \phi \cdot (A-B) }
    = - 4\pi^2 \gam\mu_{BA} \ ,
\end{equation}
which is independent of $k$. By enforcing gauge invariance the Ward
identity restricts the mixing of the bilinears.
Eq. \ref{diagramtwo} shows that, without this restriction, 
there is unwanted mixing of the Landau
gauge bilinear with the lower dimensional operator $A_\mu$.
Because of this, a continuum operator such as $\tQ_2$ will mix
with the lattice transcription of $\sbar \g\mu \pl A_\mu d$.
As mentioned at the beginning of this section, 
such mixing cannot be corrected using perturbation theory; 
it must be eliminated by a non-perturbative constraint. 
One must also worry about mixing with
operators containing two or three factors of $A_\mu$.
Thus, Landau-gauge operators require several non-perturbative
subtractions, which makes them unattractive for use in
numerical calculations of penguin diagrams.

Barring this caveat, the mixing of unsmeared Landau gauge
operators with the penguin operator
is identical to that for gauge-invariant operators. This
follows since the equations of motion for the external fermion line
annihilate any contribution from the longitudinal part of the
gluon propagator. 

For smeared Landau gauge operators additional work is
required, but, given the problems with gauge noninvariant operators,
we have not done the calculation.


\newsection{LOWER DIMENSION OPERATORS}
\label{smixing}

We close the paper by considering the mixing of SU(3) octet operators
with lower dimension operators.
We begin by summarizing the results of Ref. \cite{cps} for continuum operators.
The diagrams of Figs. \ref{lowerdimfig} and \ref{penguinfig} lead to mixing
of the $d=6$ operators listed in Eq. \ref{burasops} with quark bilinears of 
lower dimension, $d'<6$. The original operators transform in the
$(8,1)$ representation of $SU(3)_L\times SU(3)_R$, are Lorentz (pseudo)scalars,
and have positive parity under CPS
(CP followed by $s\leftrightarrow d$ interchange).
The only lower dimension bilinear with these properties is
\begin{equation}
\co_\SUB = \sbar \g\mu \pl \stackrel{\longleftrightarrow}{D_\mu} d \ ,\quad
\stackrel{\longleftrightarrow}{D_\mu} \equiv
(\stackrel{\longrightarrow}{D_\mu} - \stackrel{\longleftarrow}{D_\mu})\  .
\end{equation}
In the continuum, one does not need to worry about this operator 
because, after using the equations of motion, it can be absorbed 
into the mass matrix by a chiral rotation of the fields.
On the lattice, however, one must subtract $\co_\SUB$ by hand
\cite{sharpekvac}. This cannot be done using perturbation theory 
because the coefficient is proportional to $1/a^2$ \cite{nonpertref}. 
One must use a non-perturbative method such as that suggested in 
Ref. \cite{cps}. This uses chiral perturbation theory, 
and relies on the fact that the positive
and negative parity parts of $\co_\SUB$ are related.
The remaining issue is whether lattice artifacts invalidate
the subtraction procedure, for example by causing mixing
with additional lower dimension operators.

For Wilson fermions the subtraction procedure is, indeed, 
invalidated by the explicit breaking of chiral symmetry in the action. 
For example, there is
mixing with the $d'=3$ bilinears $\sbar d$ and $\sbar \g5 d$,
with no relationship between the two coefficients \cite{nonpertref}.
Staggered fermions, on the other hand, do have a remnant chiral symmetry on the
lattice, and it turns out that this is sufficient to allow the method
of Ref. \cite{cps} to be employed.
Part of the argument has been given in Refs. 
\cite{toolkit,wius,epsilonpus,book};
the remainder is given here.
The argument only applies, however,
if the original four-fermion operators are gauge-invariant,
and we assume this henceforth.

One point has been glossed over in this discussion.
We have assumed that the operators produced by mixing have definite
transformation properties under chiral $SU(3)$,
as is true in perturbation theory.
We must include non-perturbative effects, however,
since the mixing is with operators of lower dimension.
In particular, chiral symmetry itself is broken by the vacuum.
Nevertheless, the assumption remains valid because 
the chiral transformation properties of the operators imply
that their matrix elements satisfy Ward Identities, and these are not
affected by the breaking of chiral symmetry.

There are two questions that must be addressed with staggered fermions.
First, if we transcribe the $(8,1)$ operators onto the lattice, 
using the rules explained in Refs. \cite{toolkit,epsilonpus,book}, 
with which lower dimension operators can they mix?
Second, are the positive and negative parity parts of the operators
that one obtains related in such a way that one may use the 
subtraction method of Ref. \cite{cps}?
The answer to the first question, as we show below, is that the
only relevant mixing is with the lattice transcription of $\co_\SUB$.
Given this, Refs. \cite{toolkit,book} have answered the second question
affirmatively. Thus in the following we discuss the first question alone.
We consider only the positive parity parts of operators, i.e.
those which contribute to the $K\to\pi$ matrix elements.
The extension to the negative parity parts is obtained
by multiplying the operators listed below by $\sfno55$.

Mixing is restricted by the properties of the operators needed
to transcribe the penguin contractions of $Q_{1-6}$ onto the lattice. 
To illustrate these properties we show two examples.
The lattice transcription for the positive parity part of $\tQ_1$ is 
\begin{equation}
\label{tq1+latt}
N_f \tQ_1^\LATT =
\left(\chibar_s \gam{\mu}\chi_d \ \chibar_u \gam\mu \chi_u \right)_{II} +
\left(\chibar_s \gam{\mu5}\chi_d \ \chibar_u \gam{\mu5} \chi_u \right)_{II}
\end{equation}
Only the penguin contraction of this operator is to be considered,
i.e. that in which $\chibar_u$ and $\chi_u$ are contracted.
For the operator $\tQ_5$ the lattice transcription involves several operators.
That giving the largest contribution is
\begin{eqnarray}
N_f \tQ_5^\LATT &=& 
-2 \left( \chibar_s\iden\chi_d \left[\chibar_s\iden\chi_s
	+\chibar_d\iden\chi_d \right] \right)_{I}
\nonumber \\ \label{tq5latt}
&&\mbox{} + 2
\left( \chibar_s\sfno55\chi_d \left[\chibar_s\sfno55\chi_s
                         +\chibar_d\sfno55\chi_d \right]\right)_{I} \ .
\end{eqnarray}
Different contractions of the two parts of the operator are to be kept:
for the first line, only those in which the quark fields 
of the second bilinear are contracted with each other;
for the second line, only those where each bilinear
is contracted with one of the external mesons.
The second contraction is not a penguin diagram, but it is crucial
that it be included in order to satisfy the Ward Identity described below.
The need for this extra term is related to the fact that all the 
bilinears in Eq. \ref{tq5latt} have even distance,
in contrast to the odd distance bilinears in Eq. \ref{tq1+latt}.

These operators, and all others which arise, have the following properties:
\begin{enumerate}
\item
They are singlets under the group of reflections and rotations
which map the hypercube into itself, the hypercubic group \cite{mandula}.
Operators produced by mixing must also be singlets.
\item
They are flavor octets. Thus the bilinears produced by mixing must have the
flavor form $\chibar_s \sfno{S}{F}\chi_d$.
\item
They have positive parity under $C_0 S$, 
where $C_0$ is the lattice charge conjugation symmetry 
(see, e.g. Ref. \cite{toolkit}), 
and $S$ is the $s \leftrightarrow d$ interchange symmetry.
The transformation rules for $C_0 S$ are
\begin{eqnarray}
\nonumber
&&[U_\mu]_{ab} \longrightarrow [U_\mu^{\dag}]_{ba} \ , \quad
[F_{\mu\nu}]_{ab} \longrightarrow - [F_{\mu\nu}]_{ba} \ , \quad
m_s \leftrightarrow m_d, \\
\label{CSeqn}
&&[\chibar_s]_a \sfno{S}{F}[\chi_d]_b \longrightarrow (-)^{\Delta}
[\chibar_s]_b
\overline{(\gamma^{\dag}_{S}\otimes\xi^{\dag}_{F})}
[\chi_d]_a \ ,
\end{eqnarray}
where $a,b$ are color indices, and $F_{\mu\nu}$ is a lattice transcription
of the gluon field strength.
The lower dimension operators must therefore have positive $C_0 S$ parity.
This is possible either if the bilinear itself has positive $C_0 S$ parity,
or if it has negative $C_0 S$ parity and is multiplied by $m_s-m_d$.
\item
They satisfy a Ward Identity relating their $K^0 \pi^0$ correlator
($C_{K\pi}$) to that between the strange scalar $\kappa=\sbar d$ 
and the vacuum ($C_\kappa$):
\begin{equation}
\label{chiralWI}
\sqrt2 (m_s+m_d) \sum_{t_K} C_{K\pi}(t_K,t_\pi) = - C_\kappa(t_\pi) \ .
\end{equation}
(This is equivalent to Eq. 8.10 of Ref. \cite{toolkit}.)
This relation holds provided that one uses the operators 
$\sfno55$ and $\iden$ to create the pseudoscalars and scalar, respectively,
and provided that $m_u=m_d$.\footnote{%
For the even distance operators such as that in Eq. \ref{tq5latt}, 
there is a subtlety concerning the allowed contractions of the operator
in $C_\kappa$ which is discussed in Ref. \cite{book}.
Here we simply assume that Eq. \ref{chiralWI} applies.}
It is the lattice analog of the 
continuum Ward Identity which follows from current algebra.
It is straightforward to show that Eq. \ref{chiralWI}
can be satisfied by lower dimension operators in two ways:
(a) if the bilinear itself has positive $C_0 S$ parity, then it must
have odd distance;
(b) if the bilinear itself has negative $C_0 S$ parity, then it must 
have even distance.
These selection rules play the role of parity in the continuum CPS symmetry.
\end{enumerate}

Verstegen has classified staggered fermion bilinears under the
hypercubic group \cite{verstegen}.\footnote{%
In applying his results one must note that his
flavor matrix $T_\mu$ corresponds to our $\xi_{\mu5}$.}
Using his results we find that the following lower dimension bilinears
have all the symmetry properties listed above and thus can be produced
by mixing
\begin{eqnarray*}
{\rm dim 3:}&&
\cb_3 = \sum_\mu \chibar_s \sfno{5}{\mu5} \chi_d \\
{\rm dim 4:}&&
\cb_{4a} =
 \sum_\mu \chibar_s \gam{\mu} \stackrel{\longleftrightarrow}{D_\mu} \chi_d \\
&&
\cb_{4b} = 
\sum_\mu \sum_{\nu\ne\mu} \partial_\mu [\chibar_s \sfno{\mu}{\mu\nu} \chi_d]\\
{\rm dim 5:}&&
\cb_{5a} = \sum_{\mu\nu\rho\sigma} \epsilon_{\mu\nu\rho\sigma}
\chibar_s \sfno{\rho\sigma}{\rho5} F_{\mu\nu} \chi_d \\
&&
\cb_{5b} = \sum_{\mu\nu\rho\sigma} \epsilon_{\mu\nu\rho\sigma}
\chibar_s \sfno{\rho\sigma}{\mu5} F_{\mu\nu} \chi_d \\
&&
\cb_{5c} = \sum_{\rho,\mu} \chibar_s  \stackrel{\longleftarrow}{D_\rho}
\sfno{5}{\mu5} \stackrel{\longrightarrow}{D_\rho} \chi_d \\
&&
\cb_{5d} = \sum_\mu \chibar_s  \stackrel{\longleftarrow}{D_\mu}
\sfno{5}{\mu5} \stackrel{\longrightarrow}{D_\mu} \chi_d \\
&&
\cb_{5e} = \sum_{\mu\nu\rho\sigma} \epsilon_{\mu\nu\rho\sigma}
\chibar_s  \stackrel{\longleftarrow}{D_\mu}
\left[\sfno{\rho\sigma}{\mu}-\sfno{\rho\sigma}{\nu}\right]
 \stackrel{\longrightarrow}{D_\nu} \chi_d \\
&&
\cb_{5f} = \sum_{\mu\nu\rho\sigma} \epsilon_{\mu\nu\rho\sigma}
\chibar_s \left(
\stackrel{\longleftarrow}{D_\mu}\stackrel{\longleftarrow}{D_\nu}
- \stackrel{\longrightarrow}{D_\nu}\stackrel{\longrightarrow}{D_\mu} \right)
\left[\sfno{\rho\sigma}{\mu5}-\sfno{\rho\sigma}{\nu5}\right] \chi_d \\
&&
\cb_{5g} = \sum_{\mu\nu\rho\sigma} \epsilon_{\mu\nu\rho\sigma} \partial_\mu
[\chibar_s \sfno{\rho\sigma}{\mu5} 
\stackrel{\longleftrightarrow}{D_\nu}\chi_d]\\
&&
\cb_{5h} = \sum_{\mu\nu\rho\sigma} \epsilon_{\mu\nu\rho\sigma}
\partial_\mu [\chibar_s \sfno{\rho\sigma}{\rho5} 
\stackrel{\longleftrightarrow}{D_\nu}\chi_d]\\
&&
\cb_{5i} = \partial^2 \sum_\mu [\chibar_s \sfno{5}{\mu5} \chi_d] \\
&&
\cb_{5j} = \sum_\mu \partial_\mu^2 
[\chibar_s \sum_\mu \sfno{5}{\mu5} \chi_d] \\
&&
\cb_{5k} = \sum_{\mu\nu\rho\sigma} \epsilon_{\mu\nu\rho\sigma}
\partial_\mu \partial_\nu [\chibar_s \sfno{\rho\sigma}{\mu}\chi_d]\\
\end{eqnarray*}
$F_{\mu\nu}$ is any lattice version of the field strength with 
appropriate transformation properties under the hypercubic group.
The covariant derivatives $D_\mu$ will be defined precisely below.
We have included total derivatives in the list because they
can contribute if one calculates $K\to\pi$ matrix elements with
non-degenerate quarks.
All the operators given above can be multiplied by 
functions of $m_s$ and $m_d$ that have positive $C_0 S$ parity.
It turns out that no lower dimension operators with
explicit factors of $m_s-m_d$
(or any function of $m_s$ and $m_d$ having negative $C_0 S$ parity) 
satisfy all the properties listed above.
This is why all the operators in the list have odd distance.

The original four-fermion operators 
(e.g. Eqs. \ref{tq1+latt} and \ref{tq5latt}) 
are staggered flavor singlets.\footnote{%
We use the expression ``staggered flavor'' to distinguish from the
continuum flavor SU(3) symmetry.}
Thus, in the continuum limit, only bilinears with
$F=I$ can be produced by mixing.
In fact, of the fourteen lower dimension operators, only $\cb_{4a}$ has $F=I$. 
It is the lattice transcription of $\co_\SUB$,
and is the operator we expect to be produced.
All the other operators are staggered flavor non-singlets,
and are allowed because
elements of the hypercubic group contain flavor transformations.
What we must determine is the conditions for which these operators,
which are artifacts of the lattice regularization,
do not contribute to matrix elements.

The first condition is that we take matrix elements
between states having the same staggered flavor, 
e.g. between two pseudo-Goldstone pions each having flavor $\xi_5$. 
At first sight this solves the problem because the matrix elements
of staggered flavor non-singlet operators should vanish, 
leaving only $\cb_{4a}$.
This is only true, however, up to corrections of $O(p_{ext} a)$,
where $p_{ext}$ is an external physical momentum. 
In effect, the matrix element picks out higher dimension parts of the
original operator, and these parts can have different staggered flavors.
What one must do is take out sufficient factors of momenta until one obtains
a flavor singlet. The flavor singlet components of all the 
non-singlet operators listed above turn out to be
of dimension 6 or higher.

To explain how this works we need some results concerning
the full staggered fermion symmetry group 
including translations \cite{goltermansmit}.
In particular, we need the representations of the zero four-momentum 
subgroup, $\cg_0$, which are summarized in Ref. \cite{toolkit}.
To discuss these representations,
it is convenient to use a different definition of bilinears. 
Our choice up to now, $\co_{SF}(y)$,
(defined in Eqs. \ref{hypercuberep} and \ref{klubergeq})
has all fields lying in a single hypercube, labeled by $y$.
This definition has the disadvantage that 
$\sum_y \co_{SF}(y)$ is not a representation of $\cg_0$,
except for operators with $\Delta=0$.

Instead, following Ref. \cite{golterman}, we proceed as follows.
Each term in $\co_{SF}(y)$ is a product $\chibar(y+A) \chi(y+B)$
multiplied by a phase, where $B=_2 A+S-F$,
the subscript indicating mod-2 arithmetic. 
We first think of this term as if both $\chibar$ and $\chi$ started
at site $y+A$, with $\chi$ then shifted to site $y+B$.
If $(S-F)_\mu \ne 0$,
then in the new operators we perform this shift symmetrically
\begin{equation}
S_\mu \chi(y+A) = \half [\chi(y+A+\mu) + \chi(y+A-\mu)] \ ,
\end{equation}
rather than picking just one of these two terms.
We do this symmetric shift in all directions for which $S-F$ is non-vanishing,
resulting, in general, in an operator contained in a $4^4$ hypercube.
To complete the definition we take the average of this enlarged operator
with that obtained by shifting $\chibar$ instead of $\chi$.
This last step gives an operator which transforms into itself under $C_0$, 
with the sign given by Eq. \ref{CSeqn}.
The resulting operator we call $\co_{SF}^\TRUE(y)$,
because $\sum_y \co_{SF}^\TRUE(y)$ is part of a true representation of $\cg_0$.
In particular, it transforms into itself under translation
in the $\mu$'th direction, with phase $(-)^{\widetilde{F}_\mu}$,
where $\widetilde F_\mu =_2 \sum_{\nu\ne\mu} F_\nu$.

We also need additional ``derivative'' operators defined as for
$\co_{SF}^\TRUE$ except that one or more of the symmetric shifts
is replaced by the antisymmetric shift
\begin{equation}
D_\mu \chi(y+A) = \half [\chi(y+A+\mu) - \chi(y+A-\mu)] \ .
\end{equation}
This corresponds to taking a derivative of the quark field,
and we write the resulting bilinear as $\chibar \sfno{S}{F} D_\mu \chi$.
Clearly $(S-F)_\mu$ must be non-zero for there to be the possibility
of an insertion of $D_\mu$. 
Unlike for $\co_{SF}^\TRUE$, we distinguish explicitly between the
derivatives acting on the quark and antiquark fields.
These must be combined
to obtain eigenstates of $C_0 S$, as in $\cb_{4a}$. 
Two properties of derivative operators are important. 
First, when summed over all $y$, they give true
representations of $\cg_0$, with the translation phase being
$(-)^{\widetilde{F}_\mu}$. Second, they transform under the
hypercubic subgroup in the same way as 
$\chibar \sfno{\mu5S}{\mu5F} \chi$.
This follows because a hypercube operator $\co_{SF}$ can be written as a 
sum of $\co_{SF}^\TRUE$, which transforms exactly as $\co_{SF}$
under the hypercubic group, and a number of derivative operators.
For example,
\begin{equation}
\label{truerepeq}
\chibar \sfno{5}{\mu5} \chi =
[\chibar \sfno{5}{\mu5} \chi]^\TRUE - \half \chibar \gam{\mu}  
\stackrel{\longleftrightarrow}{D_\mu} \chi\ .
\end{equation}
This shows the decomposition of the original hypercube operators
into true representations of $\cg_0$.

We now return to the list of lower dimension operators $\cb_i$.
First we complete their definition. Some of them contain covariant
derivatives in directions for which $S-F=0$, e.g. $\cb_{5c}$.
In these cases the covariant derivative is defined by 
\[
D_\mu \chi(y+A) = \frac14\left[
\chi(j+A+2\widehat\mu) -\chi(j+A-2\widehat\mu) \right] \ .
\]
Similarly the total derivatives are defined by shifting two lattice units,
which avoids mixing with staggered flavor symmetries.
Next we note that, using results such as Eq. \ref{truerepeq},
we can rewrite all the operators so that they become true representations 
of $\cg_0$ when summed over $y$. In fact, the list already includes
all the derivative operators of dimension less than 6 that can appear.
For example, turning $\cb_3$ into a true representation using Eq. 
\ref{truerepeq}, we find the derivative operator $\cb_{4a}$.
Thus, with no loss of generality, we can take all the bilinears
in the list to already be true representations.

We can now make more precise the statement that the matrix elements
of the flavor non-singlet operators vanish up to momentum corrections.
Since $\widetilde{F}=0$ iff $F=0$, 
all such operators have a non-trivial translation phase in at least
one direction. Thus their matrix elements vanish at $p=0$, where $p$ is
the total momentum inserted by the operator.
In order to obtain, say, a $K\to\pi$ matrix element at $p=0$, 
the $K$ and $\pi$ must both have the same staggered flavor 
(since otherwise $p\approx C\pi$ for some non-zero hypercube vector $C$),
and the $K$ and $\pi$ must be degenerate
(for one can then create the states at infinite positive and negative
time, respectively, so that the operator can be averaged over all positions).
For such a matrix element only $\cb_{4a}$ contributes and the
argument is complete.
In a general lattice matrix element, however, 
there will be a small physical momentum inserted by the operator, e.g. 
if the $K$ and $\pi$ are not degenerate.
In this case the flavor non-singlet operators can have non-vanishing
matrix elements.

To evaluate these contributions we use two identities concerning
matrix elements when $(-)^{\widetilde{F}_\mu}=-1$.
The first is valid when $(S-F)_\mu\ne0$,
\begin{equation}
\label{nonzeromoma}
\vev{(\chibar \sfno{S}{F}\chi)^\TRUE (p)} = \tan(p_\mu/2)^2
\vev{\half
\chibar \sfno{\mu5S}{\mu5F} \stackrel{\longleftrightarrow}{D_\mu} \chi}
 \ ,
\end{equation}
the second for $(S-F)_\mu=0$
\begin{equation}
\label{nonzeromomb}
\vev{(\chibar \sfno{S}{F}\chi)^\TRUE (p)} = i \tan(p_\mu/2)
\vev{\chibar \sfno{\mu5S}{\mu5F} \chi} \ .
\end{equation}
There is no implied sum over $\mu$ in either equation.
The operators are defined on a single hypercube,
and the matrix element is between external states which insert momentum $p$.
For small $p$, the factors of $i\tan(p_\mu/2)$ can be replaced
by $-\partial_\mu/2$ acting on the operators.
These equations remain valid even if there are covariant derivatives
or field strengths in the operators.
Since $(-)^{\widetilde{\mu5F}_\mu}=+1$,
the operators on the right-hand-sides have trivial translation
phases in the $\mu$'th direction.
By using the equations sequentially in all directions in which 
the translation phase is non-trivial, one can find the flavor singlet
component in a given bilinear.

We now apply these equations to the bilinears $\cb_i$. 
Each application raises the dimension of the operator by either 1 or 3, 
depending on whether $(S-F)_\mu=0$ or 1.
Recall that all the bilinears $\cb_i$ are now taken to be true
representations of $\cg_0$. We work our way up from the $d'=3$ operators.
\begin{enumerate}
\item[{\em d'=3}]
There is only one such operator: $\cb_{3}$.
It can be replaced by
$\frac14 \partial^2 \chibar_s \gam{\mu}
\stackrel{\longleftrightarrow}{D_\mu} \chi_d$,
using Eq. \ref{nonzeromoma}.
This is a dimension 6 operator which, furthermore, vanishes by the
equations of motion if the quark masses are set to zero.
\item[{\em d'=4}]
The only flavor non-singlet dimension 4 operator can be written
\begin{equation}
\cb_{4b}= \sum_\mu \partial_\mu \cb_\mu\ ,\quad 
\cb_\mu=\sum_{\nu\ne\mu} \chibar_s \sfno{\mu}{\mu\nu} \chi_d \ .
\end{equation}
One application of each of Eqs. \ref{nonzeromoma} and \ref{nonzeromomb} 
shows that $\cb_\mu$ acts as a dimension 7 operator in matrix elements, 
so that $\cb_{4b}$ acts as a dimension 8 operator and can be ignored.
\item[{\em d'=5}]
Dimension 5 operators are converted
into flavor-singlets of dimension 6 or higher using Eqs.
\ref{nonzeromoma} and \ref{nonzeromomb}.
All dimension 6 operators that occur will appear in the mixing calculations
at some order. In sect. \ref{spenguin} we calculated this mixing to one-loop,
and found that the only staggered flavor singlet operators that appear
can be rewritten as four-fermion operators.
\end{enumerate}
We conclude that all the lower dimension operators except $\cb_{4a}$
have vanishing matrix elements even if non-zero physical momentum is inserted.

We close by noting that the term proportional to $I_a$ in Eq. \ref{pampzero}
is an explicit example of mixing with lower dimension operators.
At one loop this does not affect the lattice transcriptions of $Q_{1-6}$, 
but when inserted in a two loop diagram, 
it gives rise to mixing with the operator $\cb_{5a}$.
This operator has $(-)^{\widetilde F_\rho}=-1$, and $(S-F)_\rho\ne0$.
Using Eq. \ref{nonzeromoma}, one finds that it
acts in matrix elements as though it has dimension 8, and thus can be ignored.

\section*{ACKNOWLEDGMENTS}

We thank Narahito Ishizuka, Yoshihisa Shizawa and Akira Ukawa for 
very helpful discussions and correspondence, 
and for comparing results prior to publication.
We also thank Guido Martinelli for useful discussions.
SS is supported in part by the DOE under contract
DE-AC05-84ER40150 and grant DE-FG09-91ER40614, 
and by an Alfred P. Sloan Fellowship.
AP thanks the Institute for Nuclear Theory at the University of Washington
for its hospitality and the Department of Energy for partial support
during the completion of this work.

\appendix
\newsection{FIERZ TRANSFORMATIONS}
\label{appfierz}

In this appendix we give the Fierz transformation rules for the
35 diagonal scalar operators. We do {\em not} include the sign due to Fermi
statistics. Since we always do two Fierz transformations, any such
sign would cancel.

The first 25 operators are of the form exemplified by $\opold V T$,
in which the spin and flavor matrices are independent.
These form a closed set under Fierz transformation, 
with the spin and flavor part of the operators being transformed
independently. The rules can thus be obtained from those for the spin
alone:
\begin{eqnarray*}
   V - A &\leftrightarrow& A - V \\
   V + A &\leftrightarrow& 2(S - P) \\
   S + P &\leftrightarrow& \half(S+P+T) \\
   T     &\leftrightarrow& \half(3S + 3P - T) \ .
\end{eqnarray*}
Thus, for example, the transformation needed in the example of
Section \ref{ssexample} is
\[
\opold VS \to \frac18 \opold{(2S-V+A-2P)}{(S+V+T+A+P)} \ .
\]

The Fierz transformation rules for the remaining 10 operators are
\begin{eqnarray*}
\opnew{(V+A)}{(V+A)}\mu &\leftrightarrow& 
\opold{(S-P)}{(S-P)} + \opnew T T - \\
2 \opnew{(V+A)}{(V-A)}\mu &\leftrightarrow& 
                         - \opold{(S-P)}{(V-A)} - \opnew T {(V+A)} \mu \\
\opnew{(V-A)}{(V-A)}\mu &\leftrightarrow& \opnew{(V-A)}{(V-A)}\mu \\
\opnew{(V-A)} T \mu     &\leftrightarrow& \opnew T {(V-A)} \mu \\
2 \opnew{(V+A)} T \mu     &\leftrightarrow& \opold{(V-A)}{(V+A)} -
                                  4 \opnew{(V-A)}{(V+A)}\mu \\
4 \opnew T T -            &\leftrightarrow& 
              - \opold{(V+A)}{(V+A)} + 4 \opnew{(V+A)}{(V+A)}\mu \\
4 \opnew T T +            &\leftrightarrow& 4 \opold{(S+P)}{(S+P)} -
                     \opold{(S+P+T)}{(S+P+T)} \\
 && \mbox{} + 4 \opnew T T + 
\ .
\end{eqnarray*}

\newsection{Corrections from Xa diagrams}
\label{appcorr}
In this appendix we use the one-loop matching coefficients for bilinears
to calculate those for the diagonal scalar four-fermion operators.
We consider only the contributions from Xa diagrams, since,
as discussed in the text, we can bring all other one-loop calculations 
into the form of an Xa diagram.
We leave out color factors and an overall $g^2/16\pi^2$,
and set $\mu=\pi/a$, so that the anomalous dimension term is absent.
Thus, in the notation of Eq. \ref{cmdefeq}, 
we are calculating the matrix $\cm^a$.

The matching coefficients for bilinears 
(defined in Eq. \ref{bilinres}) are given in PS.
They depend on both the spin and flavor of the bilinear,
and this is codified in the labels given to the $c_i$.
For example, the correction for the bilinear 
$\gam{\nu5}$ is labeled $c_{AS}$. 
Since $c_{AS}$ does not depend on the index $\nu$, 
the correction for the operator $\opold AS$ is diagonal.
In the notation of Eq. \ref{corrnotation}
\begin{equation}
\label{appas}
\cm^a(\opold AS) = 2 c_{AS} \opold AS \ .
\end{equation}
The factor of two comes from the fact that both bilinears are corrected.

An important feature of the one-loop corrections for bilinears is that they 
commute with multiplication by $\sfno55$.
Thus if one simultaneously changes
the spin and flavor of the operators according to 
$S\leftrightarrow P$, $V\leftrightarrow A$, $T\leftrightarrow T$, 
the correction coefficients remain unchanged. 
This means, for example, that $c_{AS}=c_{VP}$.
As in sect. \ref{snotation}, we define $\cp$ to be the
operation of multiplying both bilinears in the four-fermion operator
by $\sfno55$ from the left.
Due to the axial symmetry, this operation commutes with multiplication
by the correction matrix $\cm^a$.
For example, applying $\cp$ to Eq. \ref{appas} we find
\begin{equation}
\label{appvp}
\cm^a(\opold VP) = 2 c_{AS} \opold VP = 2 c_{VP} \opold VP \ .
\end{equation}

The simple diagonal form of the correction in Eqs. \ref{appas}
and \ref{appvp}
applies for those operators whose flavor or spin is either $S$ or $P$.
In this case, the corrections do not break the flavor or spin symmetries.
For other operators, e.g. $\opold VV$, these symmetries are broken.
This is because the bilinears in these operators consist of parts
of different distances.
The bilinears in $\opold VV$ can
have $\Delta=2$ (e.g. $\sfno \mu \nu$ with $\mu\ne\nu$) or $\Delta=0$
(e.g. $\sfno \mu \mu$), and these are corrected with different 
coefficients ($c_{VV2}$ and $c_{VV0}$ respectively). 
The net result is that, in terms of the operator basis that we use, 
the corrections are
\begin{eqnarray}
\cm^a(\opold VV) &=& 2 c_{VV2} \opold VV+2(c_{VV0}-c_{VV2}) \opnew VV\mu \\
\cm^a(\opnew VV\mu) &=& 2 c_{VV0} \opnew VV\mu \ .
\end{eqnarray}
Similarly, we find
\begin{eqnarray}
\cm^a(\opold VA) &=& 2 c_{VA2} \opold VA + 2(c_{VA4}-c_{VA2}) \opnew VA\mu \\
\cm^a(\opnew VA\mu) &=& 2 c_{VA4} \opnew VA\mu \ .
\end{eqnarray}

The remaining results for operators containing vector bilinears are
\begin{eqnarray}
\label{appvteqn}
\cm^a(\opold VT) &=& (c_{VT1}+c_{VT3}) \opold VT
                     +  (c_{VT1}-c_{VT3}) \opnew VT\mu \\
\label{appvtmueqn}
\cm^a(\opnew VT\mu) &=& (c_{VT1}-c_{VT3}) \opold VT
                        +  (c_{VT1}+c_{VT3}) \opnew VT\mu \ .
\end{eqnarray}
The results for operators containing axial bilinears can be obtained
by applying $\cp$ to the above results.

Finally, we consider tensor operators.
The results for $\opold TV$ and $\opnew TV\mu$ are as in Eqs. 
\ref{appvteqn} and \ref{appvtmueqn}, with $V$ and $T$ interchanged.
The relevant coefficients are thus $c_{TV1}$ and $c_{TV3}$.
Results for $\opold TA$ can be obtained using $\cp$.
The remaining results are
\begin{eqnarray}
\nonumber
\cm^a(\opold TT ) &=& 
   (c_{TT0}-c_{TT4}) \opnew TT- + (c_{TT0}+c_{TT4}-2c_{TT2}) \opnew TT+ \\
&& \mbox{} \quad + 2 c_{TT2} \opold TT  \\
\cm^a(\opnew TT-) &=& 
   (c_{TT0}+c_{TT4}) \opnew TT- + (c_{TT0}-c_{TT4}) \opnew TT+ \\
\cm^a(\opnew TT+) &=& 
   (c_{TT0}-c_{TT4}) \opnew TT- + (c_{TT0}+c_{TT4}) \opnew TT+ \ .
\end{eqnarray}

%

\def\MPA#1#2#3{{Mod. Phys. Lett.} {\bf A#1} (#2) #3}
\def\PRL#1#2#3{{Phys. Rev. Lett.} {\bf #1} (#2) #3 }
\def\PRD#1#2#3{{Phys. Rev.} {\bf D#1} (#2) #3}
\def\PLB#1#2#3{{Phys. Lett.} {\bf #1B} (#2) #3}
\def\NPB#1#2#3{{Nucl. Phys.} {\bf B#1} (#2) #3}
\def\NPBPS#1#2#3{{Nucl. Phys.} {\bf B ({Proc. Suppl.}){#1}} (#2) #3}
\def\RMP#1#2#3{{Rev. Mod. Phys.} {\bf #1} (#2) #3}
\def\NC#1#2#3{{Nuovo Cimento } {\bf #1} (#2) #3}
\def\PREP#1#2#3{{Phys. Rep.} {\bf #1} (#2) #3}
\def\ZEITC#1#2#3{{Z. Phys.} {\bf C#1} (#2) #3}

\def\etal{{\em et al}}
\def\brookhaven{Proc. {``Lattice Gauge Theory '86''}, 
             Brookhaven, USA, Sept. 1986, NATO Series B: Physics Vol. 159}
\def\ringberg#1{Proceedings of the Ringberg Workshop, 
	{\sl ``Hadronic Matrix Elements and Weak Decays''}, 
        Ringberg, Germany, 1988, edited by A. Buras \etal,
	\NPBPS{7A}{1989}{#1}}
\def\seillac#1{
  Proceedings of the International Symposium on Lattice Field Theory,
  {\sl ``Field theory on the Lattice''},
  Seillac, France, 1987, edited by A. Billoire \etal,
  \NPBPS{4}{1988}{#1}}
\def\fermilab#1{
  Proceedings of the International Symposium on Lattice Field Theory,
  {\sl ``LATTICE 88''}
  Fermilab, USA, 1988, edited by A.S. Kronfeld and P.B. Mackenzie,
  \NPBPS{9}{1989}{#1}}
\def\capri#1{
  Proceedings of the International Symposium on Lattice Field Theory,
  {\sl ``LATTICE 89''},
  Capri, Italy, 1989, edited by N. Cabibbo \etal, 
  \NPBPS{17}{1990}{#1}}
\def\talla#1{
  Proceedings of the International Symposium on Lattice Field Theory,
  {\sl ``LATTICE 90''},
  Tallahassee, Florida, USA, 1990, edited by U. M. Heller \etal,
  \NPBPS{20}{1991}{#1}}
\def\tsukuba#1{
  Proceedings of the International Symposium on Lattice Field Theory,
  {\sl ``LATTICE 91''},
  Tsukuba, Japan, 1991, edited by M. Fukugita \etal,
  \NPBPS{26}{1992}{#1}}
\def\amsterdam#1{
  Proceedings of the International Symposium on Lattice Field Theory,
  {\sl ``LATTICE 92''},
  Amsterdam, The Netherlands, 1992, edited by J. Smit and P. Van Baal,
  \NPBPS{30}{1993}{#1}}


\begin{thebibliography}{99}
\bibitem{lusignoli}
	{M.Lusignoli, L. Maiani, G. Martinelli and L. Reina, 
	\NPB{369}{1992}{139}}
\bibitem{buras}
	{A. Buras and P. Weisz, \NPB{333}{1990}{69}}
\bibitem{bkprl}
	{G. Kilcup, S. Sharpe, R. Gupta and A. Patel, \PRL{64}{1990}{25}}
\bibitem{sharpelat91}
	{S. Sharpe, R. Gupta and G. Kilcup, \tsukuba{197}}
\bibitem{fukugitalat91}
	{M. Fukugita, N. Ishizuka, H. Mino, M. Okawa and A. Ukawa,
	\tsukuba{265}}
\bibitem{ishiprl93}
	{N. Ishizuka, M. Fukugita, H. Mino, M. Okawa, Y. Shizawa and A. Ukawa,
	 \amsterdam{415}; \PRL{71}{1993}{24}}
\bibitem{epsilonpus}
	{S. Sharpe, R. Gupta, G. Guralnik, G. Kilcup and A. Patel,
 	 \PLB{192}{1987}{149}}
\bibitem{cps}
	{C. Bernard, T. Draper, A. Soni, D. Politzer, and M. Wise,
	\PRD{32}{1985}{2343}}
\bibitem{nonpertref}
	{M. Bochicchio, L. Maiani, G. Martinelli, G. Rossi and M. Testa
	\NPB{262}{1985}{331}}
\bibitem{cpskpipi}
	{C. Bernard, T. Draper, G. Hockney and A. Soni,
	\seillac{483}}
\bibitem{sigmawilson}
	{G. Curci, E. Franco, L. Maiani and G. Martinelli, 
	 \PLB{202}{1988}{363}}
\bibitem{toolkit}
	{G. Kilcup and S. Sharpe, \NPB{283}{1987}{493}}
\bibitem{wius}
	{S. Sharpe, A. Patel, R. Gupta, G. Guralnik and G. Kilcup,
	\NPB{286}{1987}{253}}
\bibitem{book}
	{S.~Sharpe, in ``Standard Model, Hadron Phenomenology 
	and Weak Decays on the Lattice'',
	Ed. G.~Martinelli, to be published by World Scientific
	[University of Washington preprint, DOE/ER/40614-5 (July 1991)]}
\bibitem{thooft}
	{G. 't Hooft and M. Veltman, \NPB{44}{1972}{189}}
\bibitem{daniel}
	{D. Daniel and S. Sheard, \NPB{302}{1988}{471}}
\bibitem{sheard}
	{S. Sheard, \NPB{314}{1989}{238}}
\bibitem{patelsharpe}
	{A. Patel and S. Sharpe, \NPB{395}{1993}{701}}
\bibitem{tsukubapert}
	{N. Ishizuka and Y. Shizawa, Tsukuba preprint UTHEP-261 (8/93)}
\bibitem{ssringberg}
	{S. Sharpe, \ringberg{255}}
\bibitem{kieu}
	{D. Daniel and T.D. Kieu, \PLB{175}{1986}{73}}
\bibitem{kluberg}
	{H. Kluberg-Stern, A. Morel, O. Napoly and B. Petersson, 
	\NPB{220}{1983}{447}}
\bibitem{smitvink}
	{J. Smit and J. Vink, \NPB{298}{1988}{557}}
\bibitem{lepagemackenzie}
	{G.P. Lepage and P. Mackenzie, \talla{173};
	 preprint FERMILAB-PUB-19/355-T (9/92); \\
	 G.P. Lepage, \tsukuba{45}}
\bibitem{newburas}
	{A. Buras, M. Jamin, M. Lautenbacher and P. Weisz, MPI-PAE-PTH-106-92,
  	(Oct 1992), hep-ph/9211304}
\bibitem{martinelli}
	{M. Ciuchini, E. Franco, G. Martinelli and L. Reina,
         LPTENS 93/11 (April 1993), hep-ph/9304257}
\bibitem{siegel}
	{W. Siegel, \PLB{84}{1979}{193}}
\bibitem{uclapert}
	{C. Bernard, T. Draper and A. Soni, \PRD{36}{1987}{3224}}
\bibitem{altarelli}
	{G. Altarelli, G. Curci, G. Martinelli and S. Petrarca,
	 \NPB{187}{1981}{461}}
\bibitem{hmks}
	{R. Gupta, G. Guralnik, G. Kilcup and S. Sharpe,
	\PRD{43}{1991}{2003}}
\bibitem{verstegen}
	{D. Verstegen, \NPB{249}{1985}{685}}
\bibitem{sharpekvac}
	{S. Sharpe, \PLB{165}{1985}{271}}
\bibitem{mandula}
	{J. Mandula, G. Zweig and J. Govaerts \NPB{228}{1983}{91}}
\bibitem{goltermansmit}
	{M.F.L. Golterman and J. Smit, \NPB{245}{1984}{61}}
\bibitem{golterman}
	{M.F.L. Golterman, \NPB{255}{1985}{328};\\ \NPB{273}{1986}{666}}
\end{thebibliography}
\end{document}